\documentclass[12pt]{article}
\usepackage{amsmath}
\usepackage{graphicx,psfrag,epsf}
\usepackage{enumerate}
\usepackage{natbib}
\usepackage{makeidx,latexsym,rotate,amssymb,url}
\usepackage{mathrsfs}
\usepackage{pdfpages}
\newcommand{\blind}{0}

\def\tr{\mbox{\rm tr}}

\def\Var{\mbox{\rm Var}}
\def\Ex{{\rm I}\!{\rm E}}
\def\TT{^\top}
\def\ms{\sigma}
\def\argmax{\mathop{\rm argmax}}

\def\SP{{\mathscr P}}
\def\SX{{\mathscr X}}

\begin{document}
\includepdf[pages={1}]{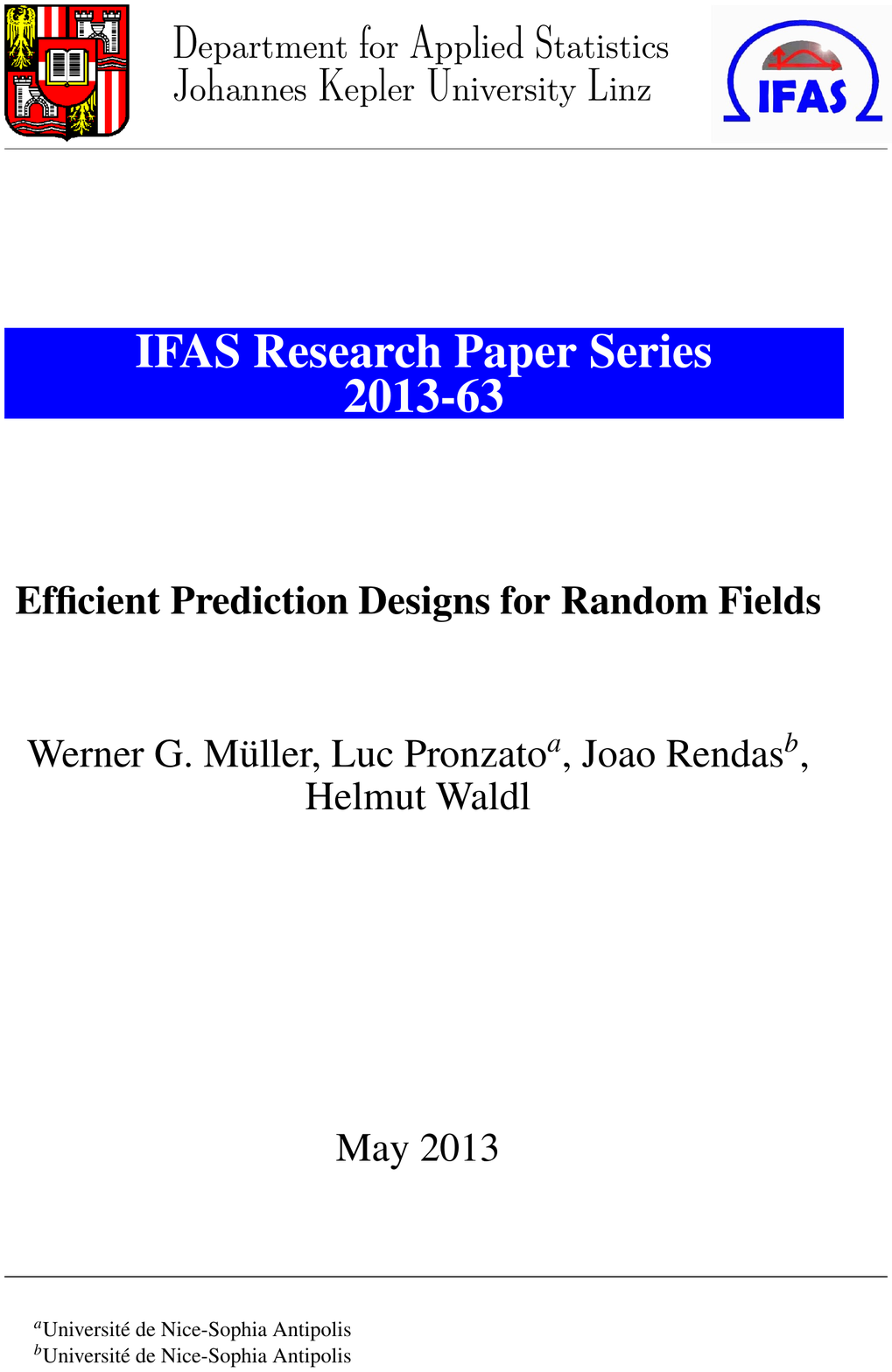}

\def\spacingset#1{\renewcommand{\baselinestretch}%
{#1}\small\normalsize} \spacingset{2}

\if0\blind

\if1\blind
{
  \bigskip
  \bigskip
  \bigskip
  \begin{center}
    {\LARGE\bf Efficient Prediction Designs for Random Fields}
\end{center}
  \medskip
} \fi

\begin{abstract}
For estimation and predictions of random fields it is increasingly acknowledged that the kriging variance may be a poor representative of true uncertainty. Experimental designs based on more elaborate criteria that are appropriate for empirical kriging  are then often non-space-filling and very costly to determine.  In this paper, we investigate the possibility of using a compound criterion inspired by an equivalence theorem type relation to build designs quasi-optimal for the empirical kriging variance, when space-filling designs become unsuitable. Two algorithms are proposed, one relying on stochastic optimization to explicitly identify the Pareto front, while the second uses the surrogate criteria as local heuristic to chose the points at which the (costly) true Empirical Kriging variance is effectively computed. We illustrate the performance of the algorithms presented on both a simple simulated example and  a real oceanographic dataset.
\end{abstract}

\noindent%
{\it Keywords:} optimal design ; pareto front; empirical kriging; gaussian process models
\vfill

\newpage
\spacingset{1.45} 
\section{Introduction}
\label{intro}

The model underlying our investigations is the  correlated scalar random
field given by

$$
Y\left( x\right) = \eta(x,\beta)+\varepsilon(x)\,.
$$

Here, $\beta$ is an unknown vector of parameters in $\mathbb{R}^p$, $\eta(\cdot,\cdot)$ a known function and the random term $\varepsilon \left( x\right)$ has zero mean, (unknown) variance $\sigma^2$ and  a parameterized  correlation structure such that $\Ex[\varepsilon \left( x\right)\varepsilon \left( x'\right)]=\sigma^2 c(x,x';\nu)$ with $\nu$ some unknown parameters. It is often assumed that the deterministic term has a linear structure, \emph{i.e.}, $\eta(x,\beta)=f\TT(x)\beta$, and that the random field $\varepsilon \left( x\right)$ is Gaussian, allowing estimation of $\beta$ and $\theta=\{\sigma^2, \nu\}$ by Maximum Likelihood. We are interested into making predictions $\hat Y(\cdot)$ of $Y(\cdot)$ at unsampled locations $x$ in a compact subset $\SX$ of $\mathbb{R}^d$ using observations $Y(x_1),\ldots,Y(x_n)$ collected at some design points $\xi=(x_1,\ldots,x_n)\subset\SX^n$. Our objective is to select $\xi$ (of given size $n$) in order to maximize the precision of the predictions $\hat Y(x)$ over $\SX$. Problems with this structure arise in such diverse areas of spatial data analysis as mining, hydrogeology, natural resource monitoring and environmental sciences, see, \emph{e.g.}, \cite{Cressie_93}, and has become the standard modeling paradigm in computer simulation experiments (cf.\ \cite{Fang+al_05,Kleijnen_09, rasmussen+w_05,  santner+al_03}), known under the designations of Gaussian Process (GP) modelling and kriging analysis.

It is conventional practice that all unknown parameters are estimated from the same data set, but clearly the classic kriging variance $\Var[\hat Y(x)]$ does not reflect the additional uncertainty resulting from the estimation of the covariance parameters; for an early discussion of this issue, see \cite{todini+f_96}.
A first-order expansion of the kriging variance for $\hat\theta$ around its true value is used in \cite{harville+j_92}, see also \cite{Abt_99} for more precise developments, leading to an explicit additive correction term to  the (normalized) kriging variance. Bootstrap solutions can be found in \cite{denhertog+al_05} and \cite{SDLunaY2003}. This  corrected kriging variance, considered in this paper, is given by

\begin{equation}\label{EK2}
MEK(\xi) = \max_{x \in \SX} \left\{ \Var[\hat Y(x)] + \tr\left\{V_\nu\,\Var[ \partial \hat Y(x) / \partial \nu]\right\} \right\}\,.
\end{equation}

The design $\xi$ that minimizes this criterion is called EK(empirical kriging)-optimal in \cite{zimmerman_06}; see also \cite{zhu+s_06} for a similar criterion. Above, $V_\nu=V_\nu(\xi,\nu)$ stands for the  covariance matrix of the estimate of the covariance parameters $\nu$ and $\hat Y(x)$ is the posterior mean of $Y(x)$ given the data at  $\xi=(x_1,\ldots,x_n)$. Note that $V_\nu$, $\Var[\hat Y(x)]$ and $\Var[\partial \hat Y(x) / \partial \nu]$ all depend on $\xi$.

In contrast to designs that simply minimize the kriging variance, EK-optimal designs are typically not space-filling, in particular for small numbers of observations. Unfortunately,  maximization of the EK-criterion  is computationally  demanding, since evaluation of (\ref{EK2}) requires the evaluation of the target function for all points in the candidate set, being unfeasible for high dimensional design spaces as it is often the case for computer experiments. It would thus be useful to have an alternative criterion that can substitute (\ref{EK2}) in the optimization procedure while still closely reflecting the actual prediction uncertainty.

The paper is organized as follows. In Section \ref{S:ET} we motivate our approach, exploiting the intimate link that should exist between the precision of predictions of the values of the field from a given dataset and the accuracy of the estimates of the process parameters based on the same observations. Section  \ref{S:PARETO} presents the actual new contributions of the paper, proposing two algorithms for identification of EK-sub-optimal designs using as surrogates two parameter estimation criteria. Two Pareto-optimal algorithms are proposed, both based on the idea of constraining the actual evaluation of MEK to points in the Pareto front of the surrogate criteria. Finally, Section  \ref{S:MUMMExample} considers the identification of Pareto-optimal designs for a spatial oceanographic  field produced by a biogeochemical mathematical model for the North Sea, and Section \ref{S:CONCL} draws conclusions on the efficiency and limitations of the approach and suggests topics for future work.

Before presenting the contributions of this paper, it is useful to consider the impact of the correction term in equation (\ref{EK2}) above, $\tr\{V_\nu\,\Var[\partial \hat Y(x) / \partial \nu]\}$: its influence diminishes as the designs get denser, which happens, for a fixed $\SX$, when the number $n$ of observations increases. Designs that minimize $\max_{x\in\SX} \Var[\hat Y(x)]$ are thus expected to resemble optimal designs for the EK-criterion when $n$ is sufficiently large.
We illustrate this on an example by comparing the behaviors of greedy procedures for the sequential construction of designs that {\em ($S_1$)} place the next design point at the current maximum of $\Var[\hat Y(\cdot)]$, or {\em ($S_2$)}  at the current maximizer of the corrected kriging variance $\Var[\hat Y(\cdot)] + \tr\{V_\nu\,\Var[\partial \hat Y(\cdot) / \partial \nu]\}$.

\begin{figure}
\begin{center}
 \includegraphics[bb= 0 35 386 376, width=.45\linewidth]{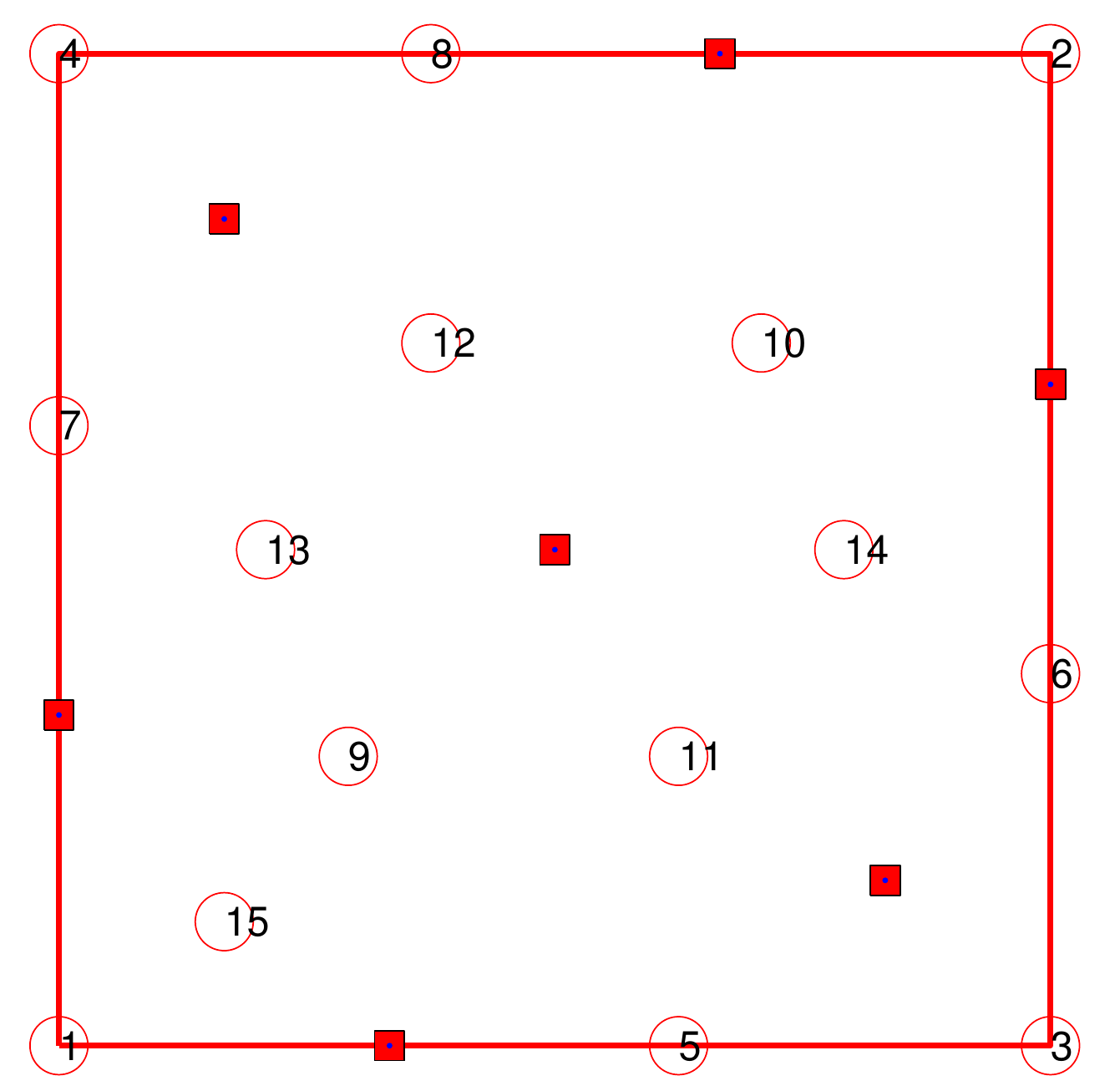} \includegraphics[bb= 0 35 386 376, width=.45\linewidth]{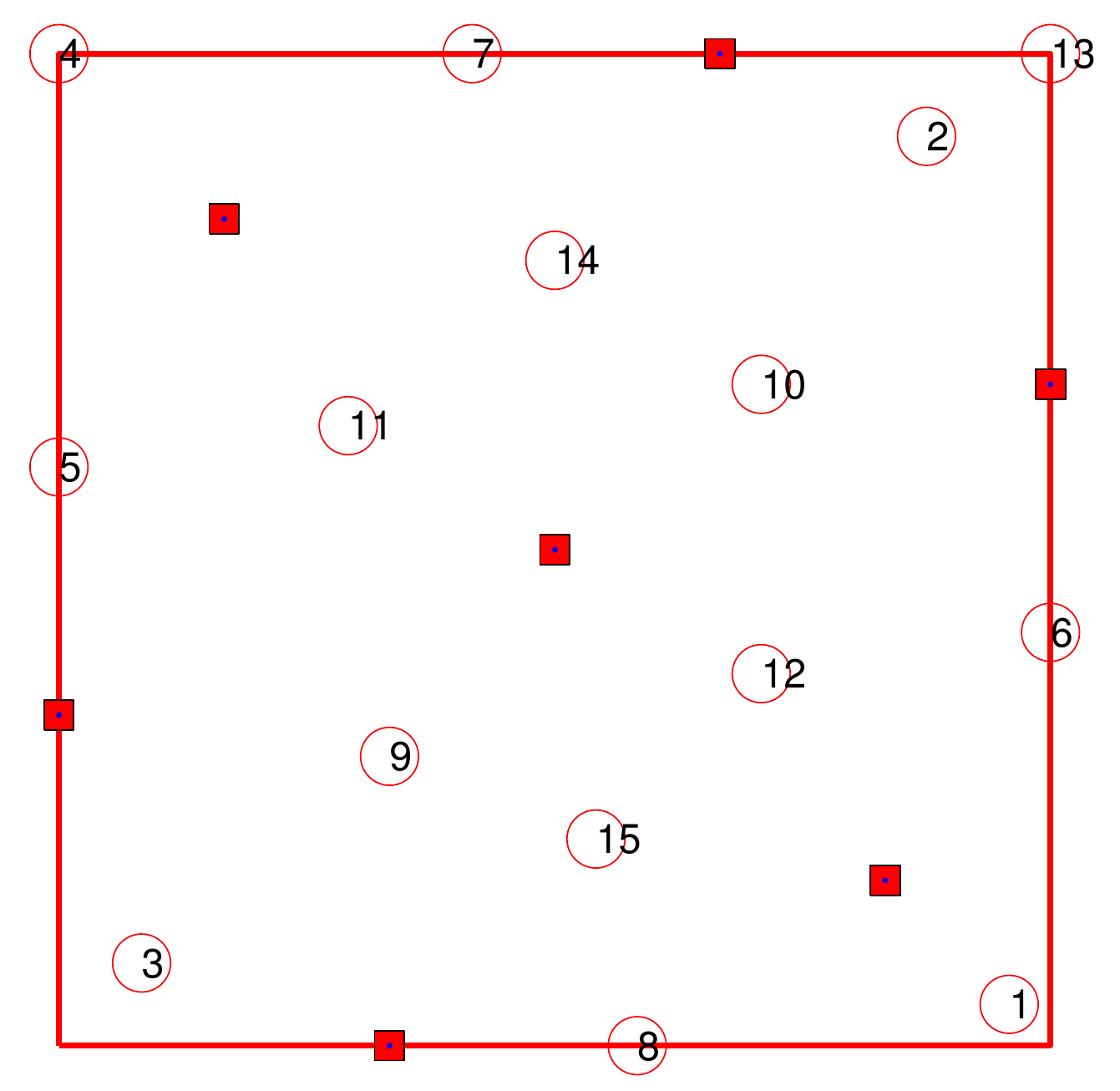}
\end{center}
\caption{\footnotesize First 15 additional points generated by the greedy strategies $S_1$ (left) and $S_2$ (right) in Example 1.}
\label{F:design_sequence15}
\end{figure}

\begin{figure}
\begin{center}
 \includegraphics[bb=0 35 476 374, width=.45\linewidth]{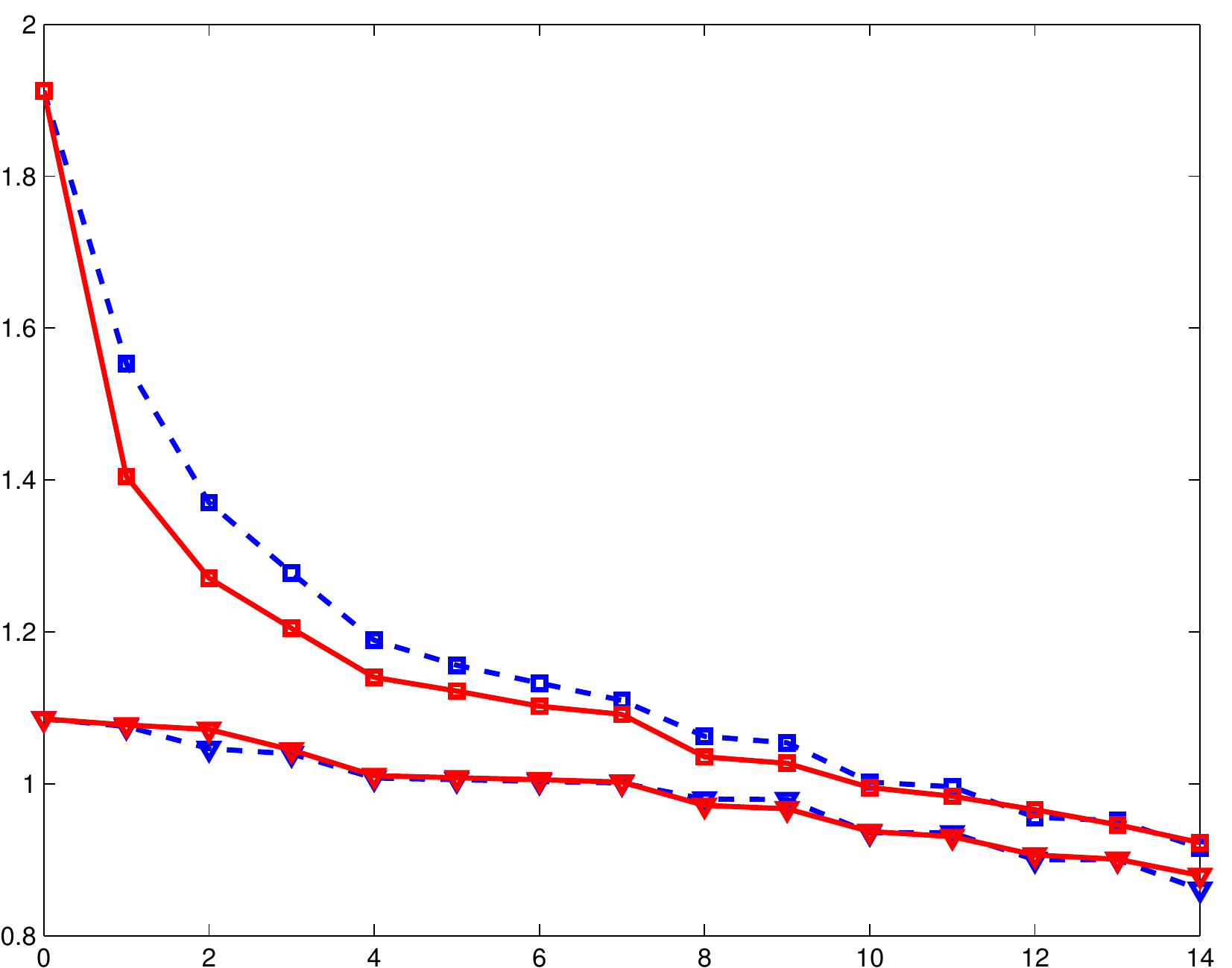}
\end{center}
\caption{\footnotesize $\max_{x\in\SX}\Var[\hat Y_k(x)]$ (triangles) and $MEK(\xi_k)$ (squares) as functions of $k$ for S$_1$ (dashed line) S$_2$ (solid line).}
\label{F:KV_EKV_sequential15}
\end{figure}

\paragraph{Example 1}
Let $\SX=[0,1]^2$, $\ms^2=1$, $c(x,x';\nu)=\exp(-\nu\|x-x'\|)$ and let  $\nu=7$. For this problem, the design
\begin{equation}\label{Lhstar}
    \xi^*_{Lh} = \left\{
\begin{footnotesize}
\begin{array}{ccccccc}
\left(
  \begin{array}{c}
    0 \\
    1/3 \\
  \end{array}
\right) &
\left(
  \begin{array}{c}
    1/6 \\
    5/6 \\
  \end{array}
\right) &
\left(
  \begin{array}{c}
    1/3 \\
    0 \\
  \end{array}
\right) &
\left(
  \begin{array}{c}
    1/2 \\
    1/2 \\
  \end{array}
\right)&
\left(
  \begin{array}{c}
    2/3 \\
    1 \\
  \end{array}
\right)&
\left(
  \begin{array}{c}
    5/6 \\
    1/6 \\
  \end{array}
\right)&
\left(
  \begin{array}{c}
    1 \\
    2/3 \\
  \end{array}
\right)
\end{array}
\end{footnotesize}
\right\} \,,
\end{equation}

\noindent plotted in Fig.~\ref{F:Lh7}-left, is simultaneously maximin and minimax optimal in the class of Latin hypercube (Lh) designs with $n=7$ points, see \cite{pronzato+m_12}.
We consider the sequential augmentation of $\xi_{Lh}^*$ with strategies $S_1$ and $S_2$ defined above. Denote by $\hat Y_k(x)$ the prediction at $x$ for the design $\xi_k=\{\xi_{Lh}^*,x_1,\ldots,x_{k}\}$, $k\geq 1$.
The design obtained by $S_1$ is space-filling, see \cite{VazquezB2011} for an analysis of its  convergence properties in terms of $\max_{x\in\SX}\Var[\hat Y_k(x)]$ as $k\rightarrow\infty$.
Figure~\ref{F:design_sequence15} shows the sequence of design points generated by the two strategies when the design space is $\{0,\,1/24,\,2/24,\ldots,23/24,\,1\}^2$.
Figure~\ref{F:KV_EKV_sequential15} shows the evolution of $\max_{x\in\SX}\Var[\hat Y_k(x)]$ (triangles) and $MEK(\xi_k)$ (squares)  given by (\ref{EK2}) as functions of $k$:  the dashed line corresponds to $S_1$ and the solid line to $S_2$.

All design points added by $S_1$ tend to fill the design space, whereas the first three points added by $S_2$ make a compromise between the precision of the prediction with $\nu$ supposed to be known and the precision of the estimation of $\nu$. However, starting with $k=4$, $S_2$ tends to be space-filling too. For $k\geq 10$ both strategies yield similar values for $\max_{x\in\SX}\Var[\hat Y_k(x)]$ and $MEK(\xi_k)$ respectively, indicating that the effect of the correcting term in $MEK(\xi_k)$ becomes negligible  as the number of observations increases.

This illustrates the fact that application of the methods presented in this paper is only justified when improvements over space-filling designs are potentially significant.
Then the impact of the correction term added to the classic kriging variance in criterion (\ref{EK2}) becomes important, which is the specific setting  addressed by this paper. Note that this may depend upon the size of the designs (smaller), the dimension of the problem (larger) and the parameter values. The problem is of practical importance whenever the cost of each observation is large, as it is the case, for instance, in geophysical applications, where it reflects both installation and maintenance of the sensing equipment.

\section{A relationship inspired by the equivalence theorem}\label{S:ET}

Intuitively, accurate predictions of a spatial field in non-observed sites requires good knowledge of the process parameters, and thus designs that optimize prediction-oriented criteria should perform well under  criteria that measure  estimation accuracy.  Such relationships are commonly exploited in the field of design of experiments  and run under the heading ``equivalence theory". They go back to the celebrated paper by Kiefer and Wolfowitz \cite{kiefer+w_60} who, by employing so-called design measures, and for parametric regression models with independent errors $\varepsilon(x)$, established the equivalence of optimal designs for two criteria of optimality, one related to parameter estimation (D-optimality), \emph{i.e.}
$$
\max_{\xi} |M_\beta(\xi)|\, ,
$$
 the other related to prediction
(G-optimality), \emph{i.e.}
$$
\min_{\xi} \max_{x \in \SX}  \Var[\hat Y(x)]\, .
$$

The analogue to G-optimality for the correlated setup considered here is the EK-criterion (\ref{EK2}) which provides a closed-form characterization of prediction uncertainty. Impacting distinct moments of the process statistical characterization, parameters $\beta$ and $\nu$, related to the trend and covariance function, respectively, have a remarkably distinct impact on the prediction error. This  motivated M\"uller and Stehl\'{\i}k \cite{mueller+s_10} to suggest the use of a convex composition of the  two corresponding D-optimality criteria as a surrogate for EK:
\begin{equation}\label{cdalpha}
J_\alpha(\xi) = \alpha \log  |M_\beta(\xi,\theta)| + (1-\alpha) \log |M_\nu(\xi,\nu)|,  \qquad \alpha\in[0,1]\,,
\end{equation}
where
$$
\left(
\begin{array}{cc}
M_\beta(\xi,\theta) \quad &0\\
0            &M_\theta(\xi,\theta) \end{array}\right) =
    \Ex \left\{
\begin{array}{ll}
-\frac{\partial^2 \log L(\beta,\theta)}{\partial\beta\partial\beta\TT}
\quad
&-\frac{\partial^2 \log L(\beta,\theta)}{\partial\beta\partial\theta\TT}\\
-\frac{\partial^2 \log L(\beta,\theta)}{\partial\theta\partial\beta\TT}
\quad &-\frac{\partial^2 \log
L(\beta,\theta)}{\partial\theta\partial\theta\TT}
\end{array} \right\} \,,
$$
with $L(\beta,\theta)$ the likelihood of $\beta$ and $\theta=(\sigma^2,\nu)$, and $M_\nu(\xi,\nu)$ in the second term of (\ref{cdalpha}) is the lower diagonal block of $M_\theta(\xi,\theta)$, with
$$
M_\theta^{-1}(\xi,\theta) = \left(
                       \begin{array}{cc}
                         a(\xi,\theta) &  b_\nu\TT(\xi,\theta)\\
                         b_\nu(\xi,\theta) &  V_\nu(\xi,\nu)\\
                       \end{array}
                     \right)\,.
$$
For the linear model $\eta(x,\beta)=f\TT(x)\beta$ simple computations lead to
$$
M_\beta(\xi,\theta) = \frac1{\sigma^2}\, \sum_{x_i \in \xi} \sum_{x_{i'}
\in \xi} f(x) [C_\nu^{-1}(\nu)]_{i,i'} f\TT(x')
$$
and
$$
\{M_\theta(\xi,\theta)\}_{ii'} = \frac12\, \tr\left\{
C_\theta^{-1}(\theta) \frac{\partial C_\theta(\theta)}{\partial\theta_i}
C_\theta^{-1}(\theta) \frac{\partial C_\theta(\theta)}{\partial\theta_{i'}}
\right\} \,,
$$
where we used the notation $\{C_\theta(\theta)\}_{ii'}=\sigma^2\, \{C_\nu(\nu)\}_{ii'}=\sigma^2\, c(x_i,x_{i'};\nu)$, $i,i'=1,\ldots,n$.
One may note that
\begin{equation}\label{Mtheta}
M_\theta(\xi,\theta) = \left(
                       \begin{array}{cc}
                         n/(2\ms^4) &  z_\nu\TT(\xi,\theta)/2\ms^2\\
                         z_\nu(\xi,\theta)/2\ms^2 &  M_\nu(\xi,\nu)\\
                       \end{array}
                     \right) \,,
\end{equation}
with
$$
\{z_\nu(\xi,\theta)\}_i = \tr\left(C_\nu^{-1} \frac{\partial C_\nu}{\partial\nu_i}\right) \ \mbox{ and } \
\{M_\nu(\xi,\nu)\}_{ij} = \frac12\, \tr\left\{ C_\nu^{-1} \frac{\partial C_\nu}{\partial\nu_i} C_\nu^{-1} \frac{\partial C_\nu}{\partial\nu_j} \right\} \,.
$$
The block $V_\nu(\xi,\nu)$ of $M_\theta^{-1}(\xi,\theta)$, which characterizes the precision of the estimation of $\nu$ and is used in (\ref{EK2}), is given by $V_\nu(\xi,\nu)=[M_\nu(\xi,\nu)- z_\nu(\xi,\theta)z\TT_\nu(\xi,\theta)/2n]^{-1}$ and does not depend on $\sigma^2$.

The reason for considering  $M_\nu(\xi,\nu)$ in the definition of $J_\alpha(\xi)$, eq. (\ref{cdalpha}), instead of the entire matrix $M_\theta(\xi,\theta)$, is that $\hat Y(x)$ is independent of $\sigma^2$, which only intervenes as a multiplicative factor in (\ref{EK2}), which thus has no influence on the optimality of a given design for the EK criterion.

The parameter $\ms^2$ is sometimes assumed to be known, and in that case $V_\nu(\xi,\nu)$ coincides with $M_\nu^{-1}(\xi,\nu)$. Assumption of knowledge about $\sigma^2$ may be motivated by estimability considerations: under the infill design framework typically not all components of $\theta=(\ms^2,\nu)$ are estimable and only some of them, or some suitable functions of them, are micro-ergodic \cite{Stein99, ZhangZ2005}; a reparame\-trization can then be used, see, \emph{e.g.}, \cite{ZhuZ2006}, with $\ms^2$ set to an arbitrary value. When both $\ms^2$ and $\nu$ are estimable, there is usually no big difference between $V_\nu(\xi,\nu)$ and $M_\nu^{-1}(\xi,\nu)$.
One may refer to \cite{mardia+m_84} for more details on these information matrices and to \cite{smirnov_05} for computationally efficient implementations for their calculation. We have preferred $M_\nu(\xi,\nu)$ over $V_\nu^{-1}(\xi,\nu)$ in the definition (\ref{cdalpha}) as it more strongly sharpens the desired balance between space-filling and nonspace-filling behaviors, see, e.g.\cite{mueller+s_10}.

Some efforts have been made to uncover quasi-equivalence relations between optimal designs for prediction and for estimation, cf.\ \cite{baldi+z:10} or \cite{mueller+al_11}. However, it was shown in \cite{mueller+al_12} that a strict equivalence between (\ref{EK2}) and (\ref{cdalpha}) does not hold, although optimal designs for  one of the criteria tend to perform well under the other, as the example below shows.

\begin{figure}
\begin{center}
 \includegraphics[bb= 0 34 517 394, width=.465\linewidth]{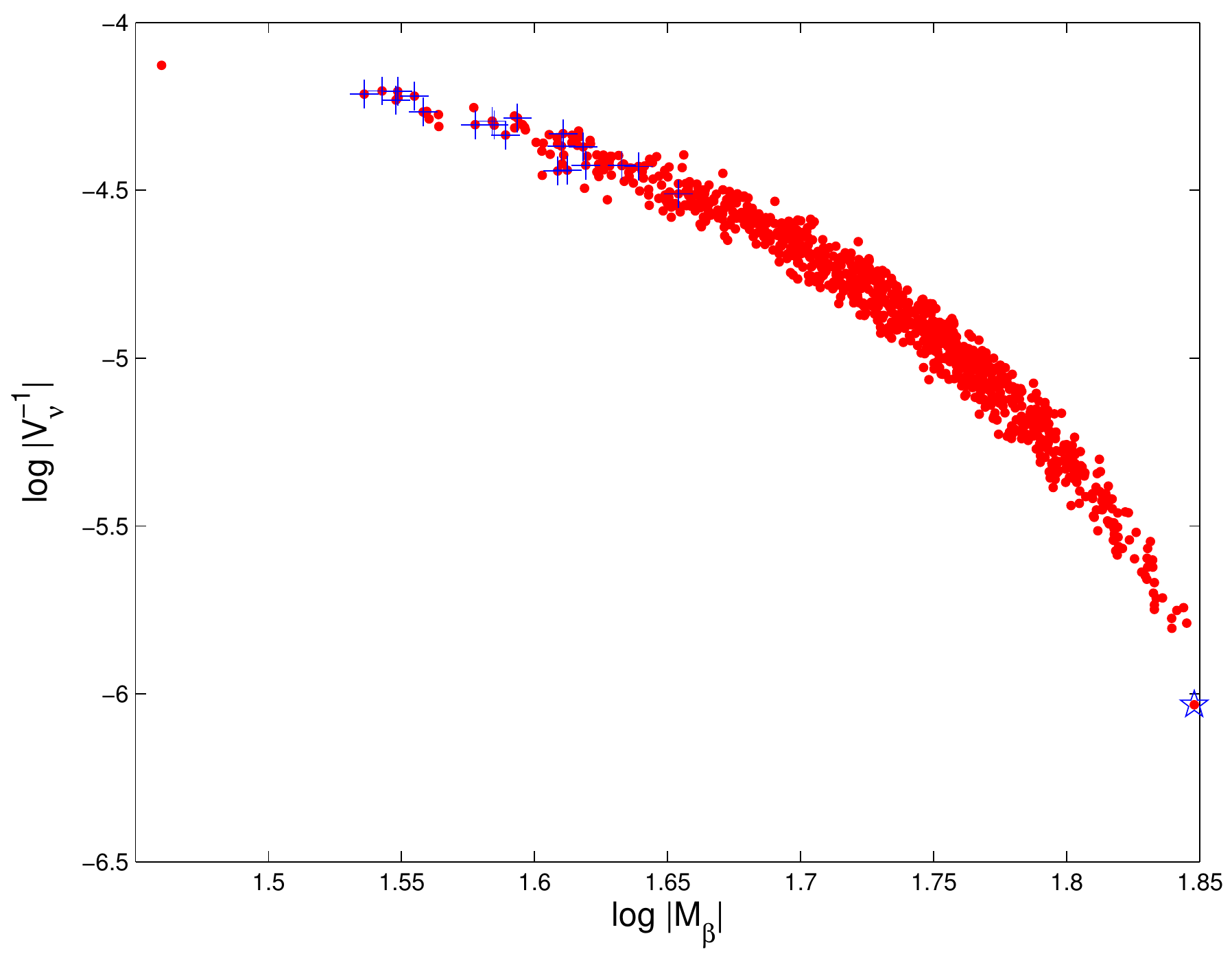} \includegraphics[bb= 0 30 499 380, width=.45\linewidth]{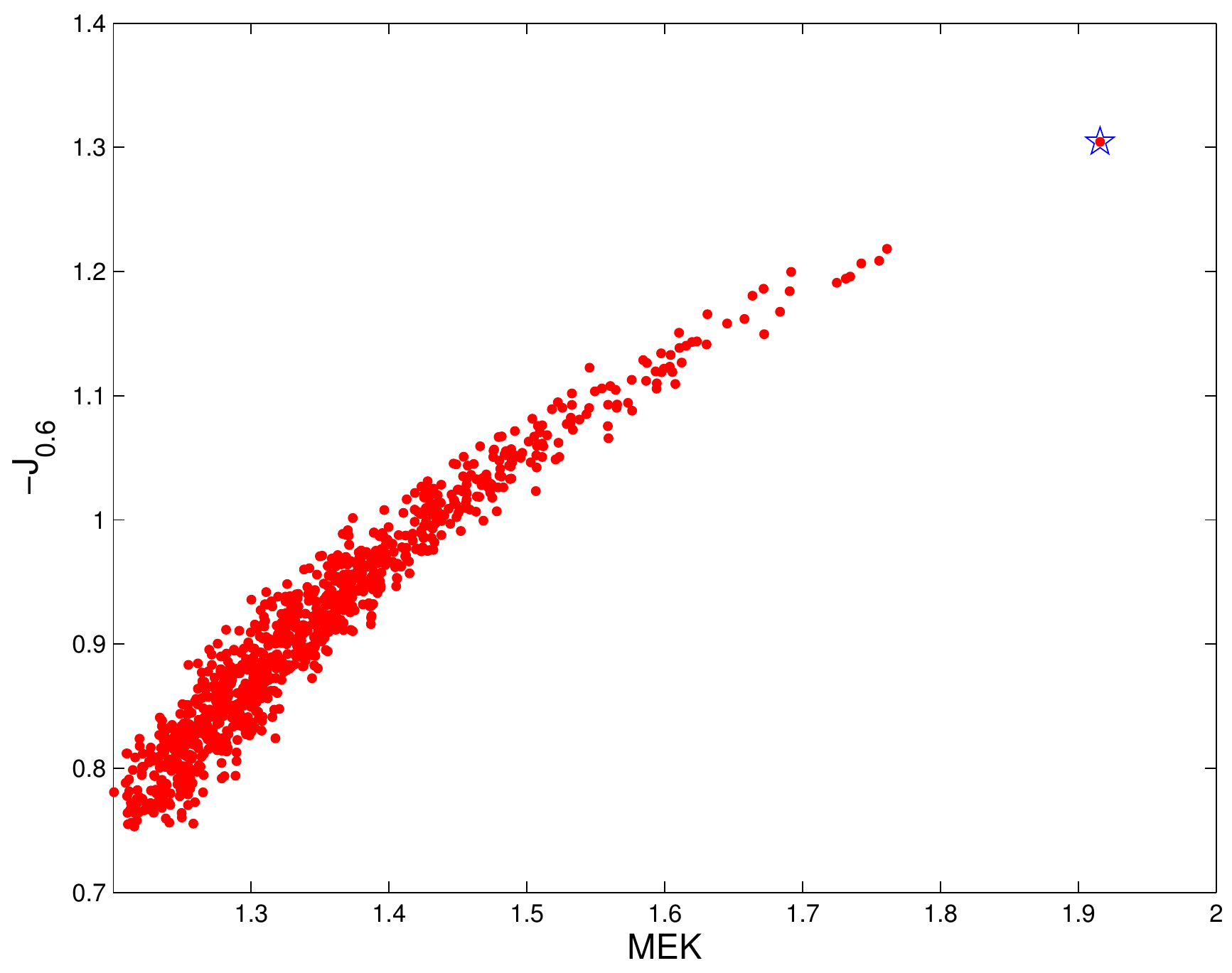}
\end{center}
\caption{\footnotesize Values of $\log |M_\nu(\xi,\nu)|$ against $\log  |M_\beta(\xi,\theta)|$ (left) and of $-J_{0.75}(\xi)$ against $MEK(\xi)$ (right) for 1,000 random Lh designs in Example 1 (the star corresponds to $\xi^*_{Lh}$).}
\label{F:M_beta-M_nu}
\end{figure}

\paragraph{Example 1 (continued)} Assume the model in Example 1, and consider 1000 i.i.d.\ random designs with $n=7$ points. Each design is a random Latin hypercube (Lh), see, \emph{e.g.}, \cite{McKayBC79}, where each component is independently perturbed by the addition of a normal random variable with zero mean and standard deviation 0.1 complemented by truncation to $[0,1]$.
Figure~\ref{F:M_beta-M_nu}-left shows the values of the two D-optimality criteria $\log  |M_\beta(\cdot,\theta)|$ and $\log |M_\nu(\cdot,\nu)|$ for these 1,000 random designs. It is quite apparent that these two criteria are antagonistic.
The blue star in the Figure corresponds to the values of the two optimality criteria for $\xi^*_{Lh}$.
As anticipated, $\xi^*_{Lh}$, which is optimal in a space-filling sense, yields a precise estimation of $\beta$ but is extremely poor for estimating $\nu$.
We also computed, for each of the random designs, the value of the EK criterion.
Figure~\ref{F:M_beta-M_nu}-right presents the values of $-J_{\alpha}(\cdot)$ for $\alpha=0.75$ against those of $MEK(\cdot)$ for the same set of designs. The first thing that we can observe is the good correlation of the two criteria for this choice of $\alpha$. Again, we note that the Lh design $\xi^*_{Lh}$ is the worst design for both criteria (they should be minimized). Points in the bottom left corner correspond to designs that are nearly simultaneously optimal for both criteria, confirming the conjecture  about the possibility of inferring EK-optimality from the two D-optimality criteria.

However, the correlation between $MEK(\xi)$ and $J_\alpha(\xi)$ observed in the example above can be much weaker for other values of $\alpha$, and the determination, without evaluating $MEK(\cdot)$, of an $\alpha^\star$ such that the maximization of $J_{\alpha^\star}(\cdot)$ yields a design close to optimality for $MEK(\cdot)$ is a difficult open problem. An expression with a structure analogous to criterion (\ref{cdalpha}) can be obtained if we search for the design that minimizes the entropy of the posterior distribution of the predicted field. The comparative analysis of the expressions of the two criteria lead to the conclusion that reasonable values of $\alpha$ must be constrained to the interval $[0.5,1]$.

\section{Pareto-optimal designs}\label{S:PARETO}

In Sect.~\ref{S:ET} we argued that finding designs $\xi$ that minimize the EK criterion (\ref{EK2}) should be intimately related to finding designs that optimize a suitable combination of the D-optimality criteria for $\beta$ and  $\nu$.
However, our ability to define a constructive experimental design method based on $J_\alpha(\cdot)$ is hampered by the lack of an efficient methodology to select  $\alpha$.

In this section we present two methods  that overcome this difficulty and that effectively lead to  design  algorithms with complexity compatible with application to real-case scenarios, as the one considered in Sect.~\ref{S:MUMMExample}.
The idea underlying both algorithms is to consider the two criteria $\log  |M_\beta(\xi,\theta)|$ and $\log |M_\nu(\xi,\nu)|$  separately, and to constrain the candidate set $\Xi$ for the minimization of (\ref{EK2}) to  the set of non-dominated designs for the corresponding multi-criteria optimization problem. The algorithms differ in the manner they approximate this non-dominated solution set. The EK criterion (\ref{EK2}) will thus play the role of a preference function for choosing designs in the reduced candidate set $\Xi$.

Other authors have addressed experimental design as a multi-criteria optimization problem, constraining the set of possible solutions to those indicated by the corresponding Pareto surface, \emph{e.g.} \cite{lu+al_11} where the author discusses its advantages over the use of scalar ``desirability functions'' and proposes methods to  chose amongst the efficient solutions of the Pareto surface. The main new contribution of our paper is the identification of two specific criteria  whose set of non-dominated solutions is a relevant (small) candidate set for  optimization of the  Empirical Kriging variance.

The set $\Xi$ of non-dominated (or Pareto-optimal) designs for the multiple objective optimization problem defined by $\log  |M_\beta(\cdot,\theta)|$ and $\log |M_\nu(\cdot,\nu)|$ is defined by
$$
\xi\in \Xi \Longleftrightarrow \forall \xi'\in\Xi\,,
\left\{
\begin{array}{l}
\log|M_\beta(\xi',\theta)| > \log|M_\beta(\xi,\theta)| \Longrightarrow \log|M_\nu(\xi',\nu)| \leq \log|M_\nu(\xi,\nu)| \\
\mbox{ and } \\
\log|M_\nu(\xi',\nu)| > \log|M_\nu(\xi,\nu)| \Longrightarrow  \log|M_\beta(\xi',\theta)| \leq \log|M_\beta(\xi,\theta)| \,.
\end{array}\right.
$$
The solid line in Fig.~\ref{F:M_beta_M_nu_2_exchange} is an example of a Pareto surface for simultaneous maximization of two criteria.

For $K$ functions $\phi_i(\cdot)$ to be maximized with respect to some variables $\xi$ and taking values that vary continuously in $K$ intervals $I_i$, the Pareto surface, or Pareto front, is in general a $(K-1)$-dimensional bounded surface included in $\bigotimes_i I_i$.
In our case, $K=2$ and the Pareto surface $\SP$ reduces to a bounded curve { --- to a finite subset of a curve when $\SX$ is finite}.
Let $P(\ell) = (C_\beta(\ell), C_\nu(\ell))$ be a parametrization of the Pareto surface.
We denote by $\{\xi\}(\ell)$ be the set of designs that map to point $P(\ell)$ in $\SP$.

In what follows we consider only designs constructed over a  finite subset $\SX_M$ of the compact design space $\SX\subset\mathbb{R}^d$, $\SX_M$ having $M$ elements. $\SX_M$ can be for instance a regular grid, with $M$ growing with $d$ like $m^d$ for some $m$, or the points of a low-discrepancy sequence, see e.g. \cite{fang+w_93}. Also, the maximization over $\SX$ in  (\ref{EK2}) will be replaced by maximization over a finite subset $\SX_{M'}$ of $\SX$ with $M'$ elements.
In general, we shall omit the index $M$ and simply write $\SX$ for $\SX_M$. Unless otherwise stated we shall take $\SX_{M'}=\SX_M$, but other choices are possible (in particular with $M'\gg M$). Also  in this paper we only consider designs without replications.

\subsection{Minimizing $MEK(\xi)$ over the set of Pareto-optimal designs}\label{S:SA-Pareto}

In general $\{\xi\}(\ell)$ is not a singleton and $MEK(\cdot)$ is not constant over this set. Moreover, the minimum of $MEK(\cdot)$ over $\SX^n$, $\xi^*$, does not generally belong to  some $\{\xi\}(\ell)$.
The minimization of $MEK(\cdot)$ over $\SX^n$ is therefore not equivalent to the minimization of $MEK(\cdot)$ over the set of Pareto-optimal designs.
However, if our belief that the two parametric estimation criteria $\log  |M_\beta(\cdot,\theta)|$ and $\log |M_\nu(\cdot,\nu)|$ yield good surrogates for the EK criterion is valid, then (\emph{i}) the variation of $MEK(\cdot)$ over each $\{\xi\}(\ell)$ should be much smaller than its variation across distant points in the Pareto surface (this fact has been checked numerically on simple examples).
$\SP$ and (\emph{ii}) the minimum of $MEK(\cdot)$ over the Pareto-optimal designs should approach the minimum of $MEK(\cdot)$ over $\SX^n$.

The method proposed in this section is based on the  identification of a finite set of Pareto optimal designs $\Xi_{\cal P}$, the final design being obtained by maximizing $MEK(\cdot)$ over this reduced set:
$$
\xi^\star_{\SP} = \argmax_{\xi\in \Xi_{\cal P}} MEK(\xi)\,.
$$

Since the Pareto surface is the set of maxima of all scalar functions monotone in each criterion, we can construct a finite set of candidate designs $\Xi_{\cal P}$   by optimizing $J_\alpha(\cdot)$ for a finite set of values of $\alpha$. However, since the maximization of $J_\alpha(\cdot)$ can only give points that belong to the convex hull of $\SP$, we may thereby miss some regions of the Pareto front.

The  optimization of $J_\alpha(\cdot)$ for fixed $\alpha$ is done using a Simulated Annealing (SA) algorithm, see \cite{Bohachevsky+al_86, citeulike:4230414, AuffrayBM_SC_2011}.
In the examples below the following  implementation of the SA algorithm has been used (remember we want to maximize $J_\alpha(\cdot)$):
\begin{description}
  \item[Step 0)] {\em Initialization}. Set initial temperature $T_0$.\\
  Draw initial design $\xi_0\propto p_0(\xi)$, $e_0 = J_\alpha(\xi_{0})$.\\
  Set current best solution $\hat \xi^\star=\xi_0$, $e^\star = J_\alpha(\hat\xi^\star)$.\\
  \vspace{-0.2cm}Set $k=0$.

  \item[Step 1)] {\em Generate candidate} $\tilde \xi_{k+1}$ by random perturbation of $\xi_{k}$: $\tilde \xi_{k+1}\propto p_{sa}(\xi|\xi_{k})$.
  \vspace{-0.2cm}
  \item[Step 2)] {\em Perform a local optimization} of $J_\alpha$ around $\tilde \xi_{k+1}$:
  \vspace{-0.2cm}$$\check\xi_{k+1} = \mbox{LocalOptimization}\left( J_\alpha(\cdot), {\tilde\xi_{k+1}}\right)\, .$$
  \item[Step 3)] \vspace{-0.2cm}{\em Update best solution}. Let $e_{k+1} = J_\alpha(\check\xi_{k+1})$.  If $e_{k+1} >e^\star$ then $\hat\xi^\star=\check\xi_{k+1}$, $e^\star=e_{k+1}$.
  \vspace{-0.9cm}\item[Step 4)] {\em Random acceptance}. If $e_{k+1} > e_{k} $ set $\xi_{k+1} = \check\xi_{k+1}$. Otherwise
  \vspace{-0.2cm}\begin{align*}
    \xi_{k+1} =\check \xi_{k}, \qquad & \mbox{with probability } p_k = \exp\left\{\frac{e_{k+1} - e_{k}}{T_k}\right\}\vspace{-0.6cm}\\
    \xi_{k+1} = \xi_{k}, e_{k+1} = e_k, \qquad & \mbox{with probability } 1-p_k
  \vspace{-3.8cm}\end{align*}
  \vspace{-1.2cm}\item[Step 5)] {\em Temperature update}. If $\xi_{k+1} = \xi_{k}$ (no change has been made in Step 3), update the temperature according to a geometric cooling scheme: $T_{k+1} = r T_{k}$.
  \vspace{-0.2cm}
  \item[Step 6)] {\em Stopping condition}. If $k=N_{max}$ stop; otherwise $k \leftarrow k+1$, return to Step 1.
\end{description}
Throughout the algorithm we keep track of the best solution found, which is eventually reported as $\xi^\star_{\SP}$. It is also expedient to start the algorithm with a space-filling design $\xi_0$ to quickly weed out the cases for which our method is obviously unnecessary.

Like most random-search algorithms, under assumptions  that are easily satisfied the SA algorithm above allows us to reach an arbitrary neighborhood (in terms of criterion value) of a global maximum of $J_\alpha(\cdot)$ in a finite number of iterations almost surely, see, \emph{e.g.}, \cite{AuffrayBM_SC_2011}. However, convergence may be slow and the risk of stopping the algorithm well before reaching some reasonable neighborhood of an optimal solution cannot be neglected.

The random perturbation $p_{sa}(\xi|\xi_{k})$ in Step 1 consists in the  replacement of two randomly chosen points $(x_i,x_j)$ of $\xi_k$ by two points uniformly drawn (without replacement) from $\SX_M\setminus\xi_k$.

In Step 2, Local Optimization$(J_\alpha(\cdot),\xi)$ is a procedure that performs iterative optimization of $J_\alpha(\cdot)$, starting from design $\xi$. Our implementation assumes that $\SX_M$ is a regular rook-type grid on which we define the clique $V_x$ of point $x\in {\SX_M}$ as the set of its NSWE (NSWE: North, South, West, East) neighbors in $\SX_M$.
\medskip

\begin{figure}
\begin{center}
 \includegraphics[bb= 0 60 480 440, width=.5\linewidth]{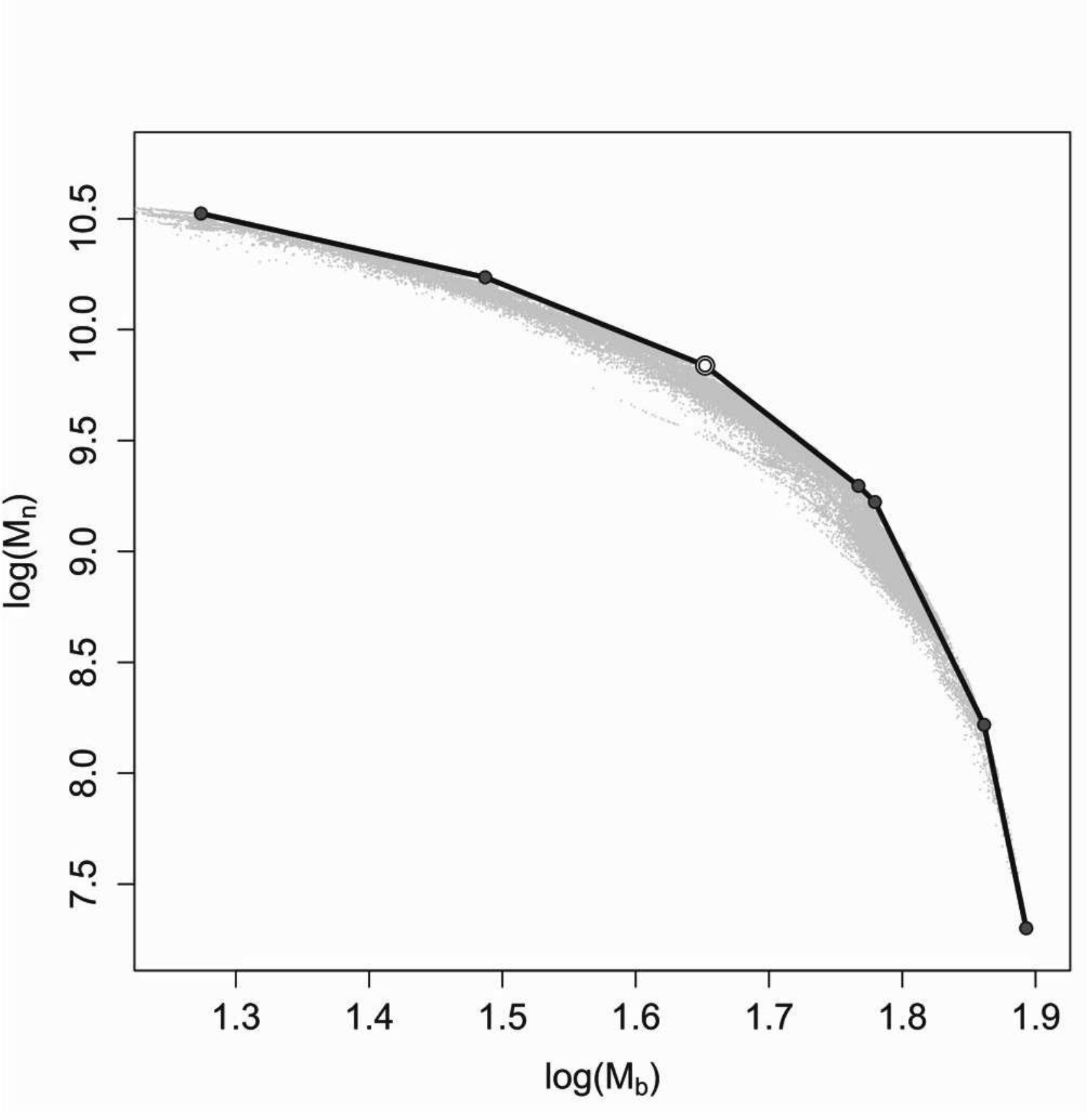}
\end{center}
\caption{\footnotesize Sampled points (in grey) for generating the Pareto surface (black); 7 points form the convex hull, the one in white being selected.} \label{fig:ParetoSimple}
\end{figure}

\noindent{\bf Local Optimization}$(J_\alpha(\cdot),\xi)$\\
\noindent Do\  \{ \\
\makebox[.25cm][l]{ } Set $\xi_0 = \xi$\\
\makebox[.25cm][l]{ } Set $J_0 = J_\alpha(\xi_0), J = J_0$.\\
\makebox[.25cm][l]{ } For all $ x_i\in\xi_0$ (scan all  points in $\xi$)\\
\makebox[1cm][l]{ }For $  x \in V_{x_i}\cap \SX_M$ (consider replacement by all points in the clique of $x_i$)\\
\makebox[1.5cm][l]{ } Set $\tilde \xi = (x_1,\ldots, x_{i-1}, x,x_{i+1},\ldots,x_d)$, $\tilde J =  J_\alpha(\tilde \xi)$ \\
\makebox[1.5cm][l]{ } If $\tilde J  > J$ set $\xi = \tilde \xi$, $J= \tilde J$ \\
\} while $ J > J_0$\\
Return($\xi_0$)

\begin{figure}
\begin{minipage}[t]{0.47\linewidth}
\includegraphics[bb= 10 10 450 450, width=\linewidth]{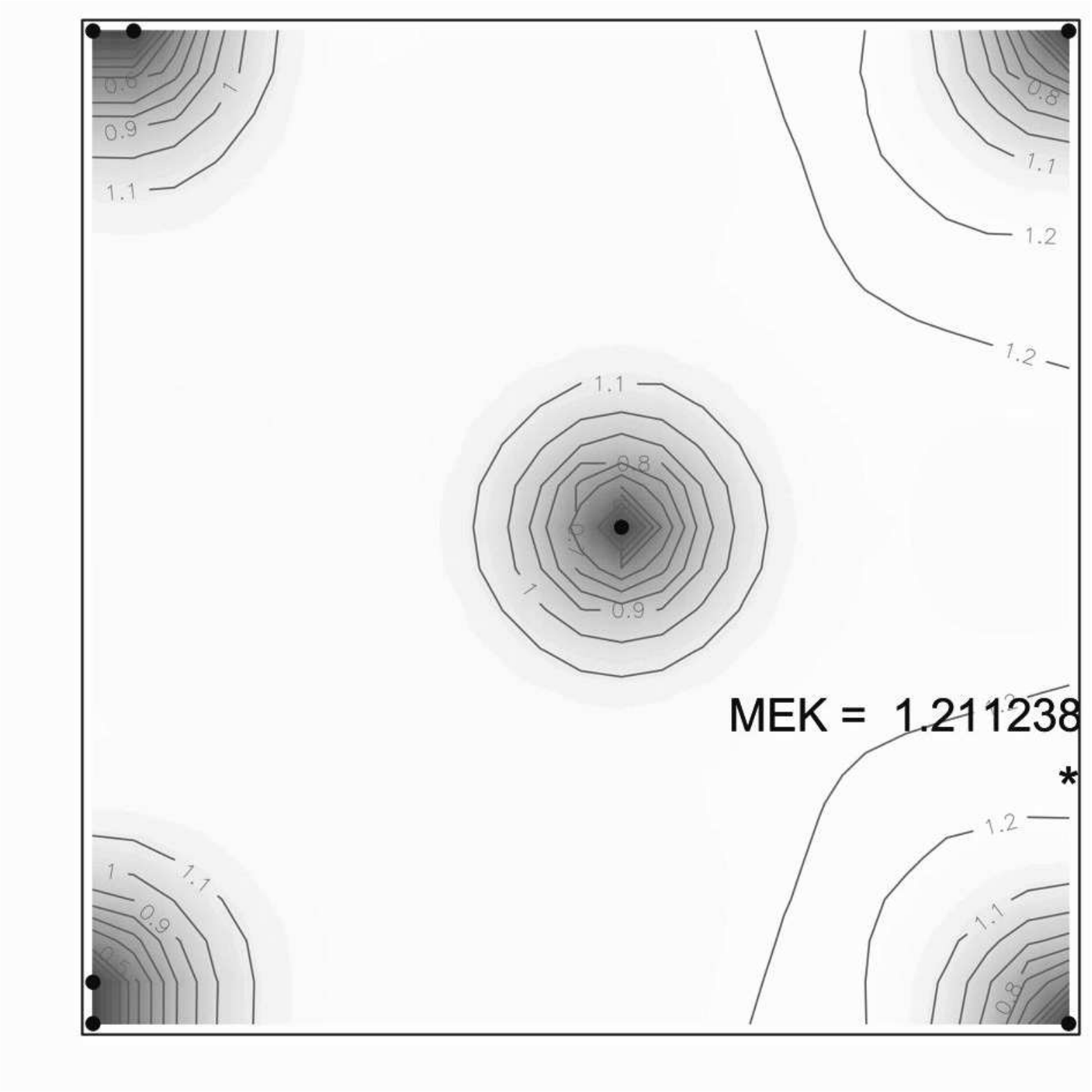}
\vspace{-1cm}
\caption{\footnotesize Corrected Kriging variance for the Pareto-optimal design found. Black dots indicate the design points. }\label{fig:BestDesignSimple}
\end{minipage}
\hfill
\begin{minipage}[t]{0.47\linewidth}
\includegraphics[bb= 10 10 450 450, width=\linewidth]{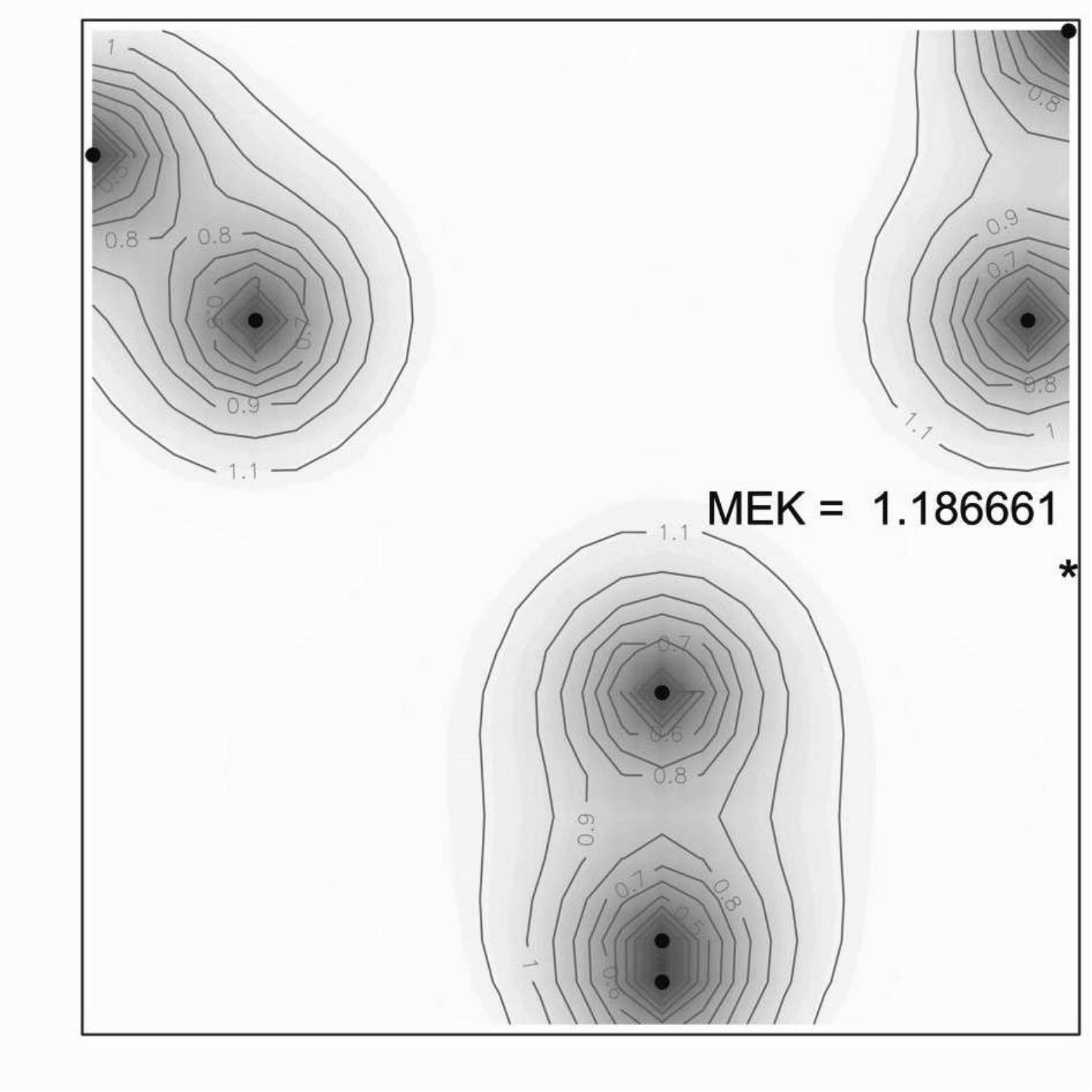}
\vspace{-1cm}
\caption{\footnotesize Corrected Kriging variance over field of analysis for the EK-optimal design. Black dots indicate the sampled points.}
\label{fig:OptimalSimple}
\end{minipage}
\end{figure}

\paragraph{Example 1 (continued)}
We illustrate now, for the process introduced in Example 1, the application of this method for finding 7-point designs for prediction over the finite design space ${\SX_M} = \{0,1/24,\ldots,23/24, 1\}^2$.

Figure \ref{fig:ParetoSimple} shows the 7 distinct values on the Pareto surface obtained by maximization of $J_\alpha$ for 11 values of $\alpha$ uniformly spread  in $[0.5,1]$. the black dots indicate the values for all designs generated during the optimization, the asterisk indicates the location of the maximum. The following parameters were used for the SA algorithm: $T_0=0.6$, $r=0.93$,  $N_{max}=5000$. Tests over a large number of executions of the SA lead to no noticeable variations of the Pareto-front in Figure \ref{fig:ParetoSimple}.

$MEK(\cdot)$ was subsequently computed for the 7 Pareto-designs  and $\xi^\star_{\SP}$ selected as the best one:
\begin{equation}\label{xi_*}
\xi^\star_{\SP} = \left\{
\begin{footnotesize}
\begin{array}{ccccccc}
\left(
\begin{array}{c}
0 \\
0 \\
\end{array}
\right) &
\left(
\begin{array}{c}
0 \\
1/24 \\
\end{array}
\right) &
\left(
\begin{array}{c}
0 \\
1 \\
\end{array}
\right) &
\left(
\begin{array}{c}
1/24 \\
1 \\
\end{array}
\right)&
\left(
\begin{array}{c}
13/24 \\
1/2 \\
\end{array}
\right)&
\left(
\begin{array}{c}
1 \\
0 \\
\end{array}
\right)&
\left(
\begin{array}{c}
1 \\
1 \\
\end{array}
\right)
\end{array}
\end{footnotesize}
\right\} \,,
\end{equation}

In Figure \ref{fig:BestDesignSimple} we present a contour plot of the corrected Kriging variance for $\xi^\star_{\SP}$. In the plot, the black dots indicate the design points, at which the variance is zero.

We also searched directly for the optimal $EK$ design $\xi^\star$ by optimizing $MEK(\cdot)$ using the SA algorithm. The much higher computational complexity of criterion evaluation imposed in this case constraining the maximum number of iterations of the Simulated Annealing algorithm to $N_{max}=2000$ . The optimal design obtained is shown in Figure \ref{fig:OptimalSimple} along with the corresponding surface of corrected Kriging variance. The effectiveness of the method can be appreciated by computing the efficiency of the Pareto-optimal design $\xi^\star_{\SP}$ with respect to the optimal design $\xi^\star$, which is in this case $EK(\xi^\star)/EK(\xi_{\SP}^\star) = 1.187/1.211 \simeq 0.98$.

\begin{figure}
\begin{center}
\includegraphics[bb= 40 50 400 365, width=.5\linewidth]{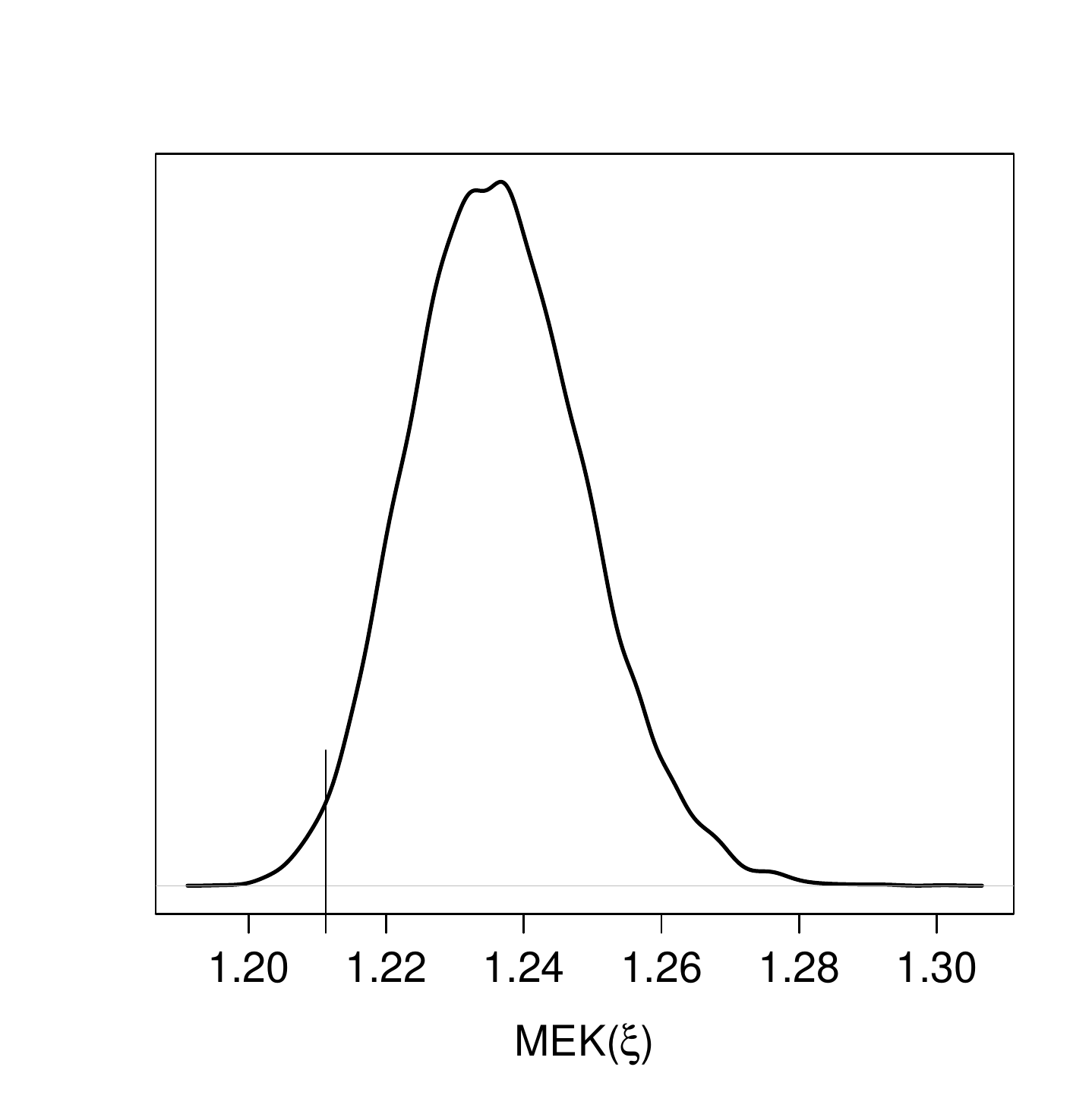}
\end{center}
\caption{\footnotesize Empirical distribution of EK-Minima for 10000 sets of randomly generated designs; vertical bar indicates our design}
\label{fig:EKdist}
\end{figure}

Notice that the construction of $\xi^\star_{\SP}$ only required 7 evaluations of the expensive criterion $MEK(\cdot)$. So for completeness, we now simulated $10000$ random sets of 7 designs $\{u_i\}_{i=1}^7$ and computed $\min_i EK(u_i)$  for each. The empirical distribution of these minima is given in Figure \ref{fig:EKdist}.
It shows that 98\% of the random designs generated with the same effort as ours lead to a corrected kriging variance larger than the one obtained using $\xi^*$.

\subsection{A simplified exchange algorithm}\label{S:simple exchange}

\begin{figure}
\begin{center}
 \includegraphics[bb= 0 30 356 356, width=.45\linewidth]{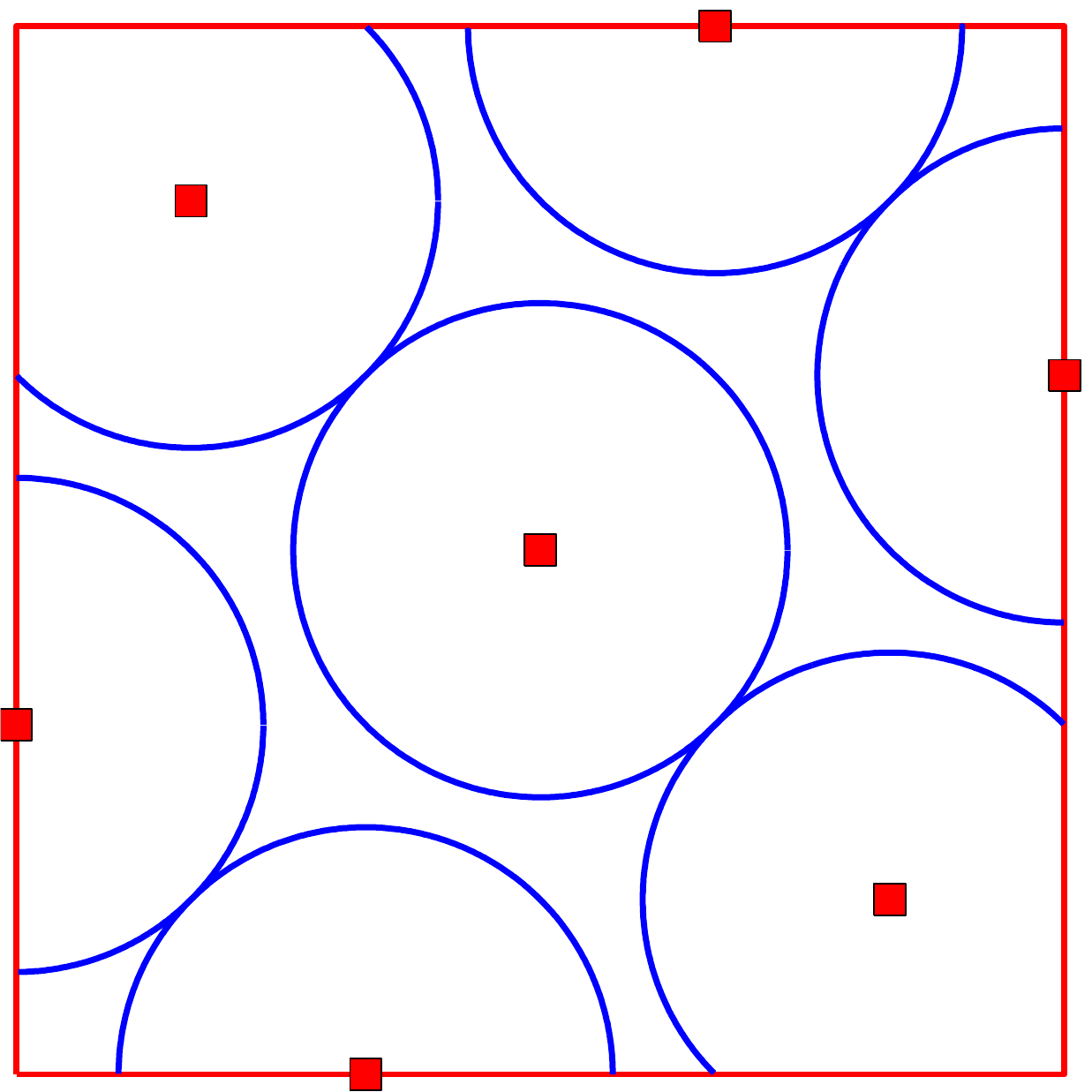} \includegraphics[bb= 0 30 356 356, width=.45\linewidth]{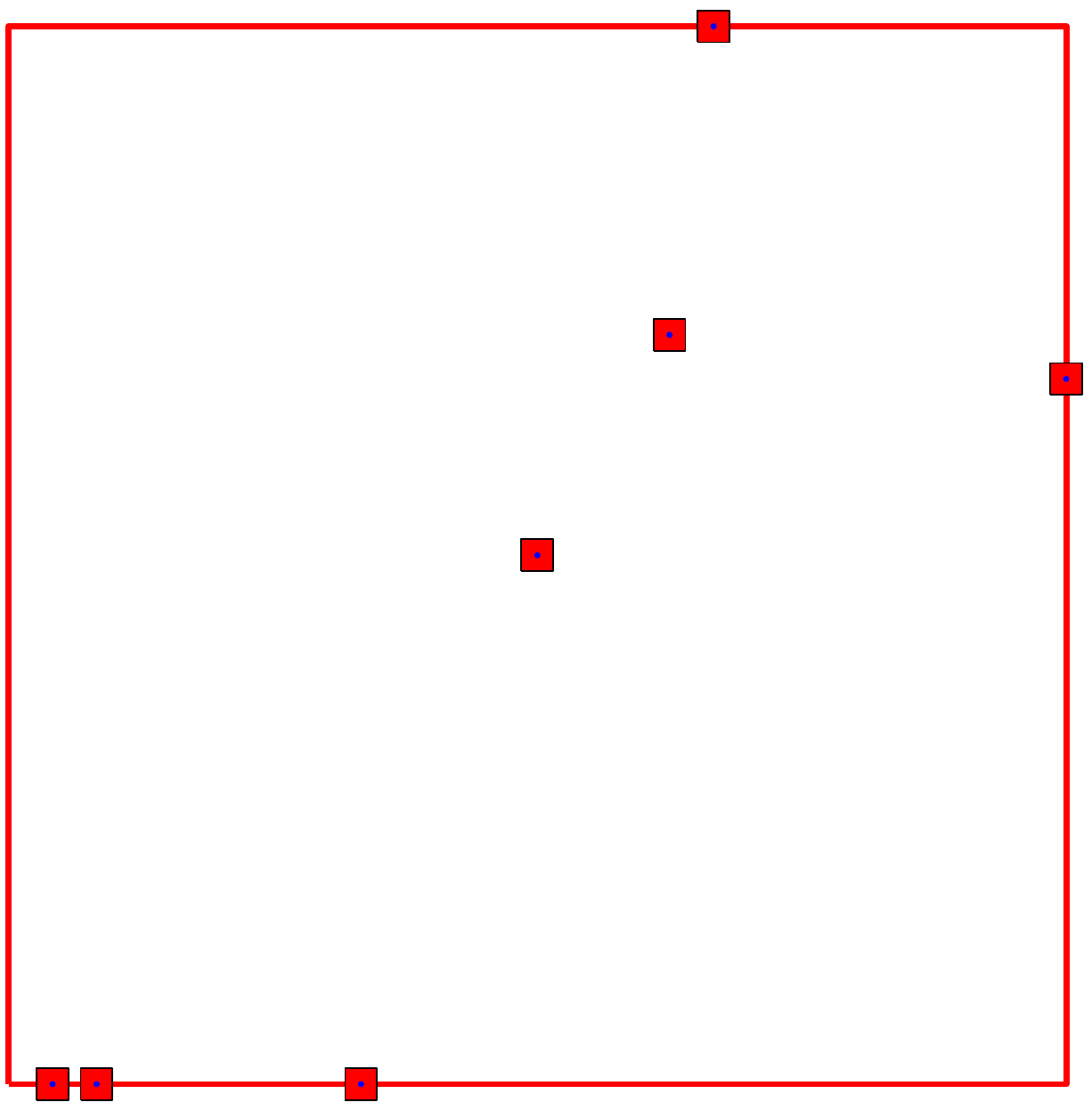}
\end{center}
\caption{\footnotesize Lh design $\xi_{Lh}^*$ (\ref{Lhstar}) (left) --- the circles have radius $\min_{x_i\neq x_j\in\xi_{Lh}^*} \|x_i-x_j\|$ --- and design $\xi_4$ (\ref{xi_4}) (right)}
\label{F:Lh7}
\end{figure}

The method proposed in this section is based on an idea suggested in \cite{pronzato+m_12}. Like the algorithm above, it makes use of the Pareto front, but, in contrast to it, is deterministic, stops after a finite number of iterations when $\SX$ is finite, and therefore cannot provide any guarantee of asymptotic convergence.
We call {\em exchange} the substitution of one point $x\in\SX$ for one point $x_i$ of the current design $\xi$.  For any given design $\xi$ with $n$ distinct points in $\SX$ there are thus $n\times (M-n)$ possible exchanges. The algorithm starts with an arbitrary design, \emph{e.g.} space-filling, and exchanges one point at a time; only exchanges corresponding to non-dominated solutions for the two criteria $\log  |M_\beta(\cdot,\theta)|$ and $\log |M_\nu(\cdot,\nu)|$ are retained for the evaluation of $MEK(\cdot)$; the best among them gives the design carried to the next iteration.
\begin{description}
  \item[Step 0)] \emph{Initialization}. Choose a space-filling design $\xi_0$ with $n$ points (\emph{e.g.}, a Lh design), compute $MEK^*_0=MEK(\xi_0)$, set $k=0$.
  \vspace{-0.3cm}\item[Step 1)] \emph{Construction of the Pareto front}. Construct the $N_k$ designs $\xi_k^i$ corresponding to all possible exchanges for $\xi_k$ and compute the associated values of $\log  |M_\beta(\xi_k^i,\theta)|$ and $\log |M_\nu(\xi_k^i,\nu)|$, $i=1,\ldots,N_k$; construct the subset $\Xi_k$ of designs $\xi_k^i$ that correspond to non-dominated solutions for $\log  |M_\beta(\cdot,\theta)|$ and $\log |M_\nu(\cdot,\nu)|$.
  \vspace{-0.3cm}\item[Step 2)] \emph{Evaluation of the EK-criterion}. Compute $MEK(\xi_k^i)$ for all $\xi_k^i$ in $\Xi_k$.
  \vspace{-0.3cm}\item[Step 3)] \emph{Design update}. If $\min_{\xi_k^i\in\Xi_k} MEK(\xi_k^i) \geq MEK^*_k$, stop; \\
  otherwise set $\xi_{k+1}=\arg \min_{\xi_k^i\in\Xi_k} MEK(\xi_k^i)$,
  $MEK^*_{k+1}=  MEK(\xi_{k+1})$, $k \leftarrow k+1$, return to step~1.
\end{description}
At step 1, $N_0=n\times (M-n)$ exchanges are considered at first iteration, but $N_k=(n-1)\times (N-n)$ for $k\geq 1$ since we do not need to consider the exchange of the same point of $\xi$ for two consecutive iterations. Also, not all $\xi_k^i$, $\log  |M_\beta(\xi_k^i,\theta)|$ and $\log |M_\nu(\xi_k^i,\nu)|$ have to be stored since the set of non-dominated solutions $\Xi_k$ can be constructed iteratively.
A further simplification is obtained by restricting $\Xi_k$ to designs that correspond to points on the convex hull of the Pareto front (which can also be constructed iteratively). A continuation of Example 1 gives an illustration.

\begin{figure}
\begin{center}
 \includegraphics[bb= 0 30 517 401, width=.55\linewidth]{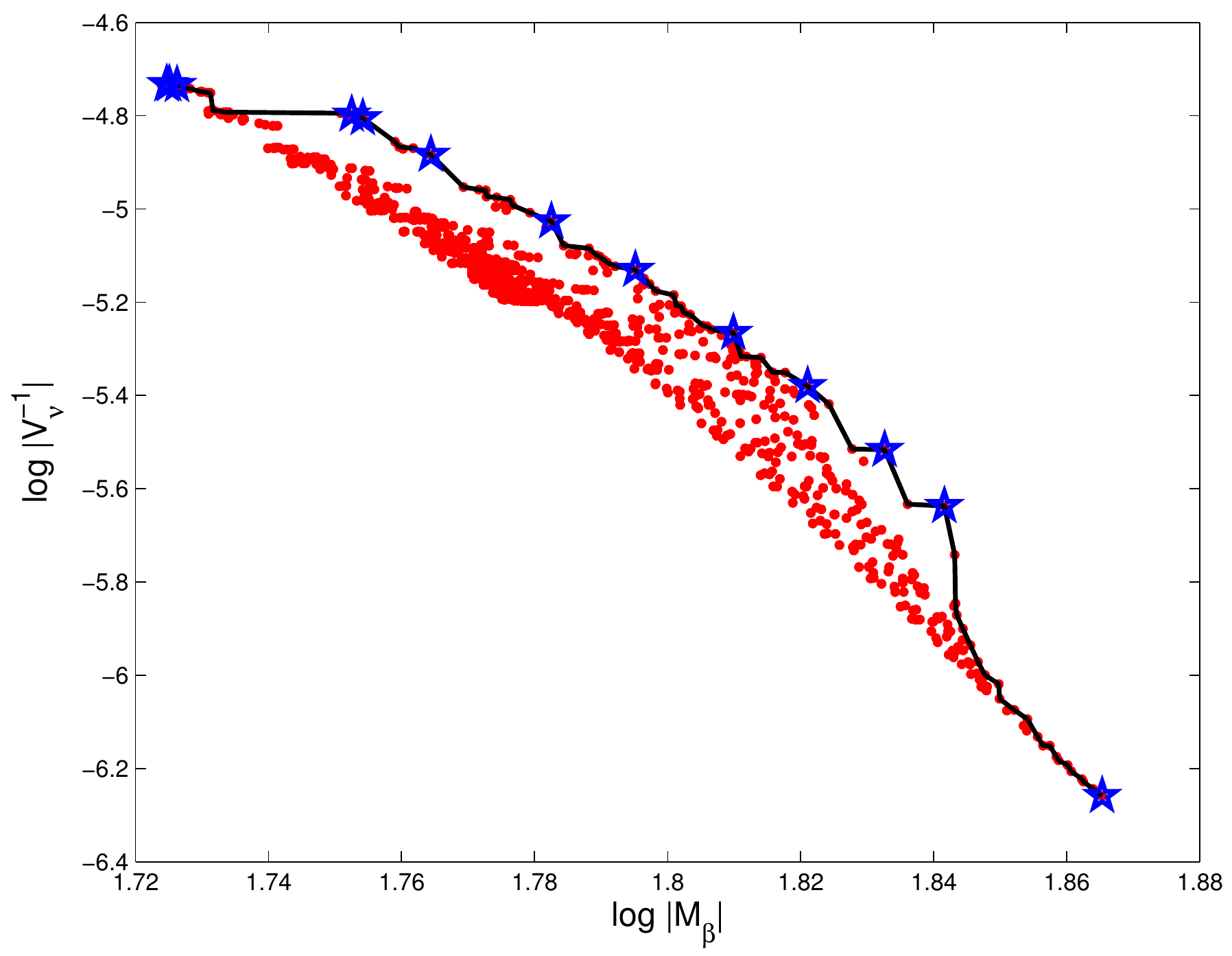}
\end{center}
\caption{\footnotesize Values of $\log |M_\nu(\xi_k^i,\nu)|$ against $\log |M_\beta(\xi_k^i,\theta)|$, $i=1,\ldots,N_0$, at iteration 1 of the simplified exchange algorithm in Example 1: the stars correspond to points on the convex hull of the Pareto front, which is indicated by the solid line.}
\label{F:M_beta_M_nu_2_exchange}
\end{figure}

\paragraph{Example 1 (continued)} We again  restrict $\SX$ to the $25\times 25$ grid of points with coordinates in the set ${\SX_M}$. Note that this set contains the design $\xi_{Lh}^*$ given by (\ref{Lhstar}), which is chosen as initial design $\xi_0$ (with $MEK(\xi_0)\simeq 1.9124$).
The algorithm above, with $\Xi_k$ given by all points on the pareto front stops after 3 iterations and returns a design with an MEK of 1.2060 requiring 967 evaluations of the EK-criterion. When $\Xi_k$ is restricted to the points on the convex hull of the Pareto front the algorithm stops after 4 iterations and returns the design
\begin{equation}\label{xi_4}
    \xi_4 = \left\{
\begin{footnotesize}
\begin{array}{ccccccc}
\left(
  \begin{array}{c}
    1/3 \\
    0 \\
  \end{array}
\right) &
\left(
  \begin{array}{c}
    0 \\
    1/3 \\
  \end{array}
\right) &
\left(
  \begin{array}{c}
    2/3 \\
    1 \\
  \end{array}
\right) &
\left(
  \begin{array}{c}
    1 \\
    2/3 \\
  \end{array}
\right)&
\left(
  \begin{array}{c}
    23/24 \\
    2/3 \\
  \end{array}
\right)&
\left(
  \begin{array}{c}
    3/8 \\
    0 \\
  \end{array}
\right)&
\left(
  \begin{array}{c}
    0 \\
    1 \\
  \end{array}
\right)
\end{array}
\end{footnotesize}
\right\} \,,
\end{equation}
see Fig.~\ref{F:Lh7}-right, with $MEK(\xi_3)\simeq 1.2080$. Figure~\ref{F:M_beta_M_nu_2_exchange} shows the values of $\log  |M_\beta(\xi_k^i,\theta)|$ and $\log |M_\nu(\xi_k^i,\nu)|$, $i=1,\ldots,N_0=4\,326$, at the first iteration of the algorithm. There are 296 non-dominated points on the Pareto front (in solid line), but only 15 points (indicated by stars) on its convex hull. The restriction of $\Xi_k$ to those points thus reduces the computational cost significantly: the EK-criterion (\ref{EK2}) is only evaluated 45 times in total when the algorithm stops. Note that although this is six times more often than the procedure of section 3.1 it gave a slight improvement of the criterion and is still considerably quicker than the simulated annealing procedure.

\section{Application to a real oceanographic field}\label{S:MUMMExample}

\begin{figure}
\begin{center}
\includegraphics[bb= 0 45 280 310, width=.47\linewidth]{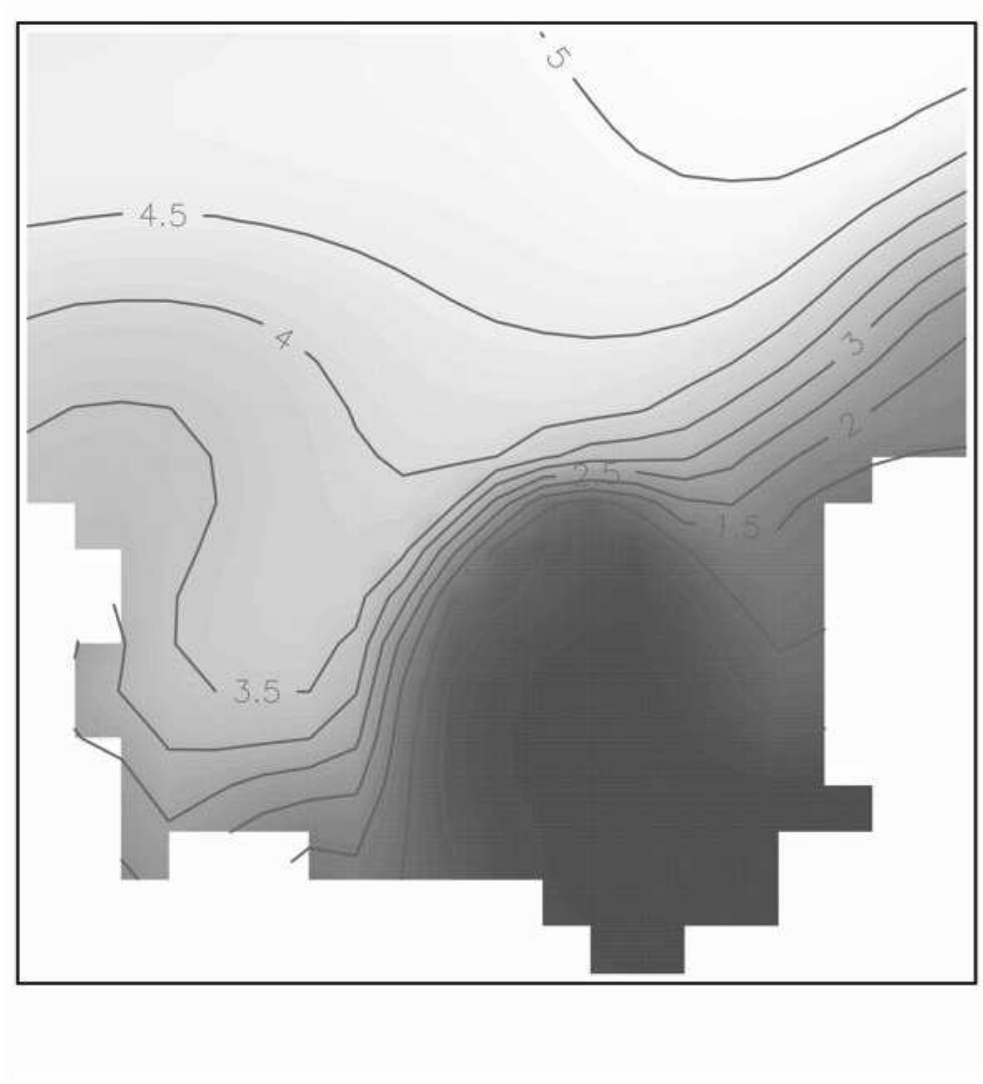}
\end{center}
\caption{\footnotesize Ammonium field over the region of interest. }\label{fig:originals_98}
\end{figure}

This section  presents the application of the design algorithm described in section \ref{S:SA-Pareto} to a real oceanographic  dataset. The data used in this study was made available through a collaboration with the institute MUMM, a department of the Royal Belgian Institute of Natural Sciences. The data is the output of the biogeochemical oceanographic model MIRO\&CO \cite{Lacroix_2007}.
MIRO\&CO- 3D is run to simulate the annual cycle of inorganic and organic carbon and nutrients, phytoplankton, bacteria and zooplankton with realistic forcing conditions. The model covers the entire water column of the Southern Bight of the North Sea, while in the study presented here we concentrate on an horizontal (sea surface) grid of $21\times 21$ points corresponding to the Belgian Coastal Zone (BCZ).

The model results from the integration of 4 modules  describing: (i) the dynamics of phytoplankton, (ii) zooplankton, (iii) bacteria and dissolved/particulate organic matter degradation and (iv) nutrient (nitrate (NO3), ammonium (NH4), phosphate (PO4) and dissolved silica (DSi)) regeneration in the water column and the sediment.
The field considered here is one of the maps of the distribution of NH4, illustrated in Figure \ref{fig:originals_98},  and our goal is to identify the 7-point design that would enable the best prediction of the NH4 field simulated by the model over the other points of the grid. This problem is representative of the design of  networks of fixed oceanography stations with limited size.

\begin{figure}
\begin{center}
 \includegraphics[bb= 0 45 482 485, width=.49\linewidth]{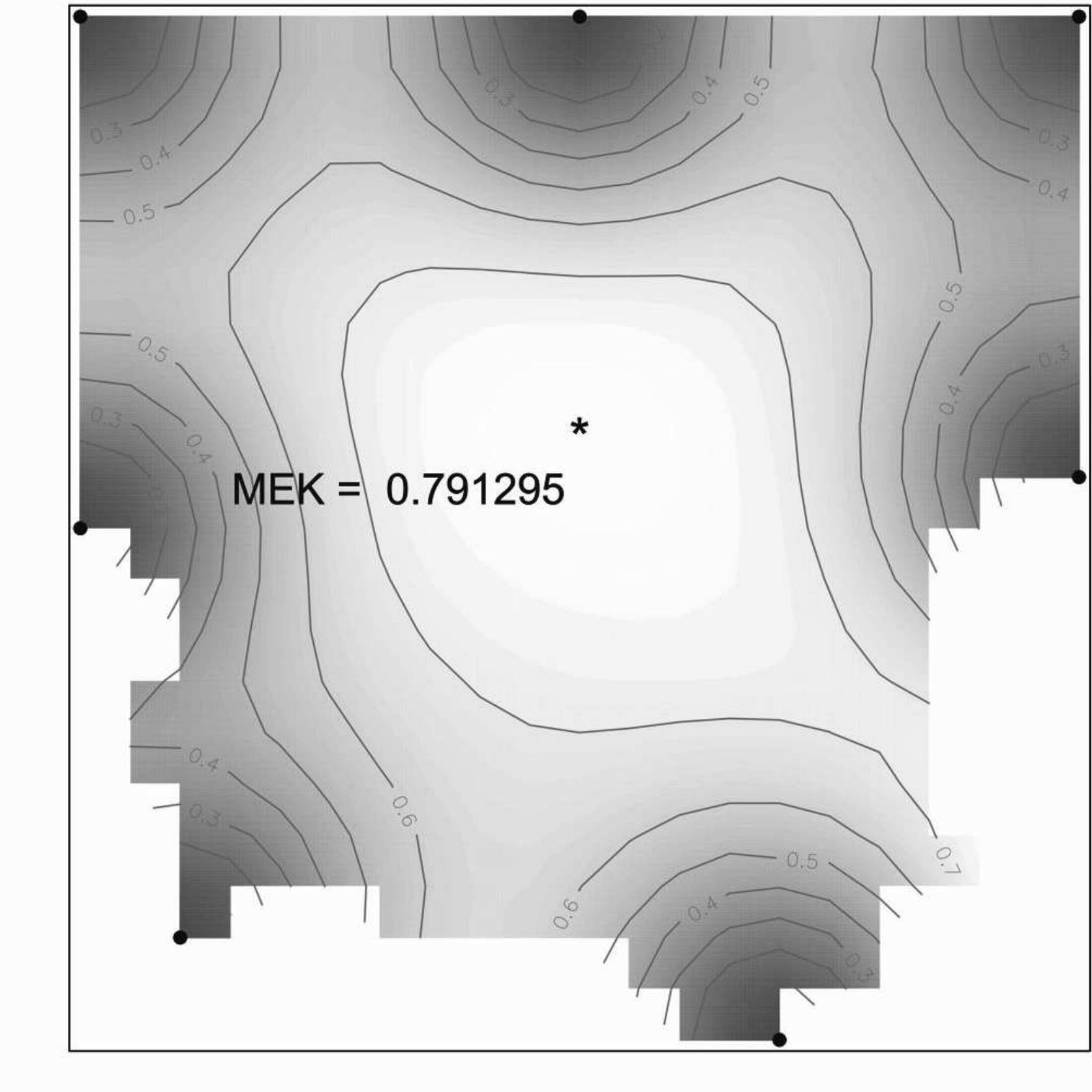}\includegraphics[bb= 0 45 482 485, width=.49\linewidth]{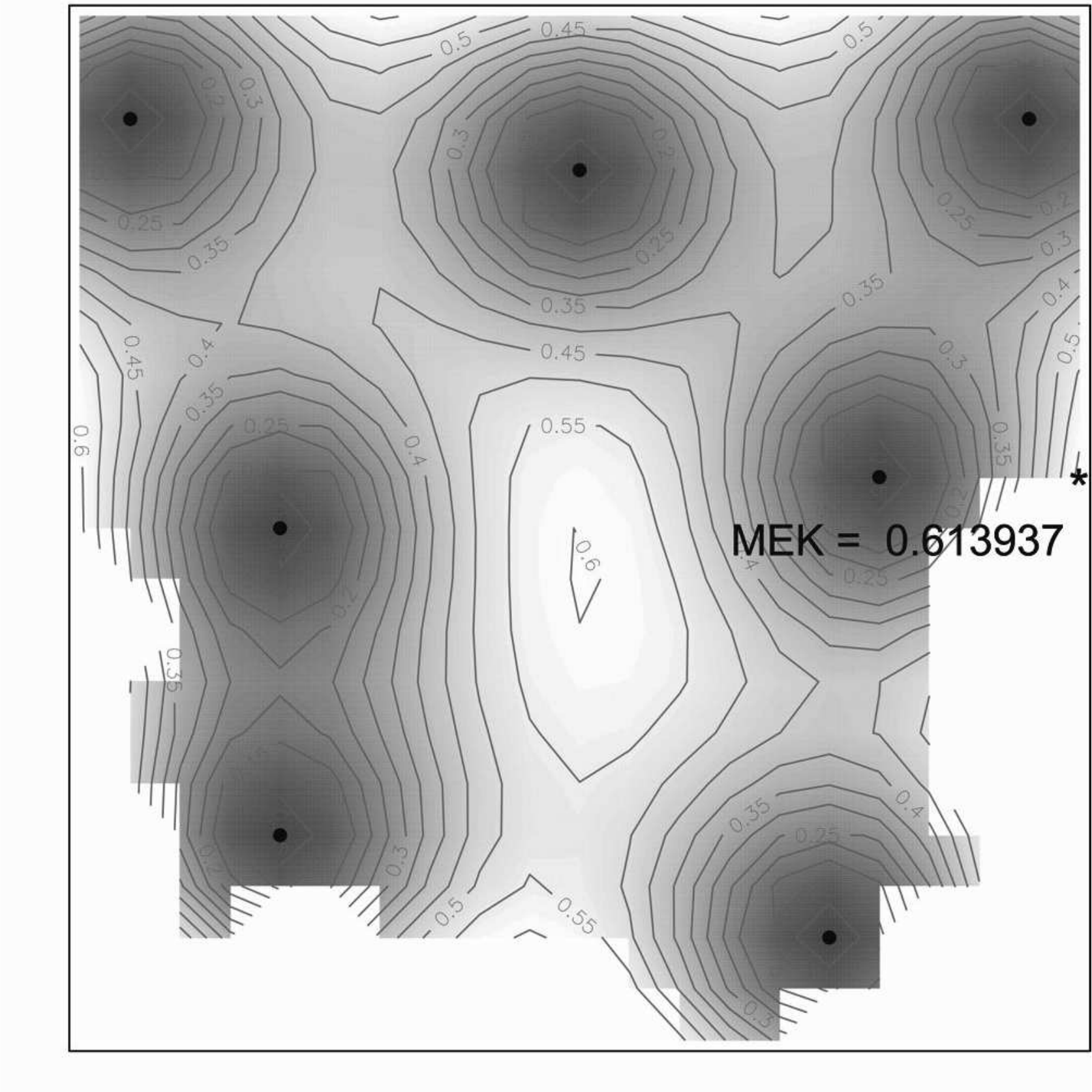}
\end{center}
\caption{\footnotesize Corrected Kriging variance over field of analysis for the best designs found. Left: Pareto-based algorithm; right: direct optimization of EK.}\label{fig:BestDesignMUMM}
\end{figure}

Since our design criteria depend on the true process characteristics, we started by
fitting a GP model to the model output. Using Maximum Likelihood, we fitted  the available data using a simple model with linear trend and Mat\'ern covariance function
\[
c(x,x^\prime,\nu) = \frac{(\left\|x-x^\prime\right\|/\rho)^{\gamma}}{\Gamma(\gamma)2^{\gamma-1}} K_{\gamma}\left( \left\|x-x^\prime\right\|/\rho\right), \qquad \nu = [\rho, \gamma]\enspace ,
\]
where $K_\nu(\cdot)$ is the modified Bessel function of the second kind of order $\nu$, obtaining
\[
\eta(x_1,x_2, \beta) = - 1.511 - 0.051 x_1-0.210 x_2\enspace ,
\]
$\sigma^2 = 0.728$, and range parameter $\rho = 2.723$. The smoothness parameter was held fixed at $\gamma = 3/2$, which gives $c(x,x^\prime,\rho)=(1+\|x-x'\|/\rho)\,\exp(-\|x-x'\|/\rho)$.

The  $7$ point Pareto-optimal design $\xi_{\SP}^\star$ for this model in the region of analysis has then been found by the method  presented in  section \ref{S:SA-Pareto}, where 6 distinct points were identified on the convex hull of the Pareto-surface. The  parameters of the SA algorithm were set as in Example 1, that was started from a random initialization.
The minimal Empirical Kriging variance was identified for $\alpha > 0.8$, indicating the importance of a good fit to the trend term in this case.

\begin{figure}
\begin{center}
 \includegraphics[bb= 0 60 369 390, width=.49\linewidth]{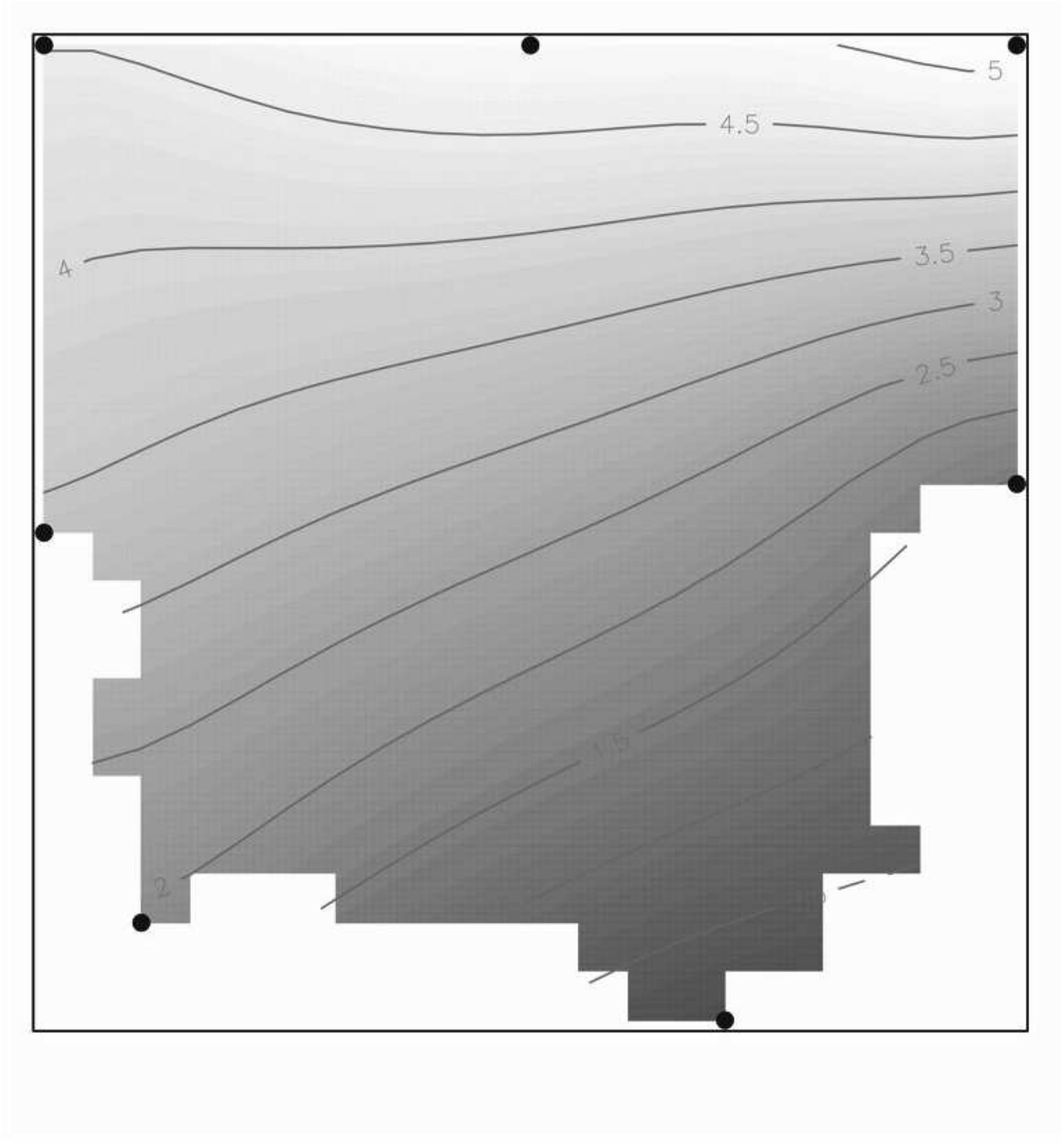}\includegraphics[bb= 0 60 369 390, width=.49\linewidth]{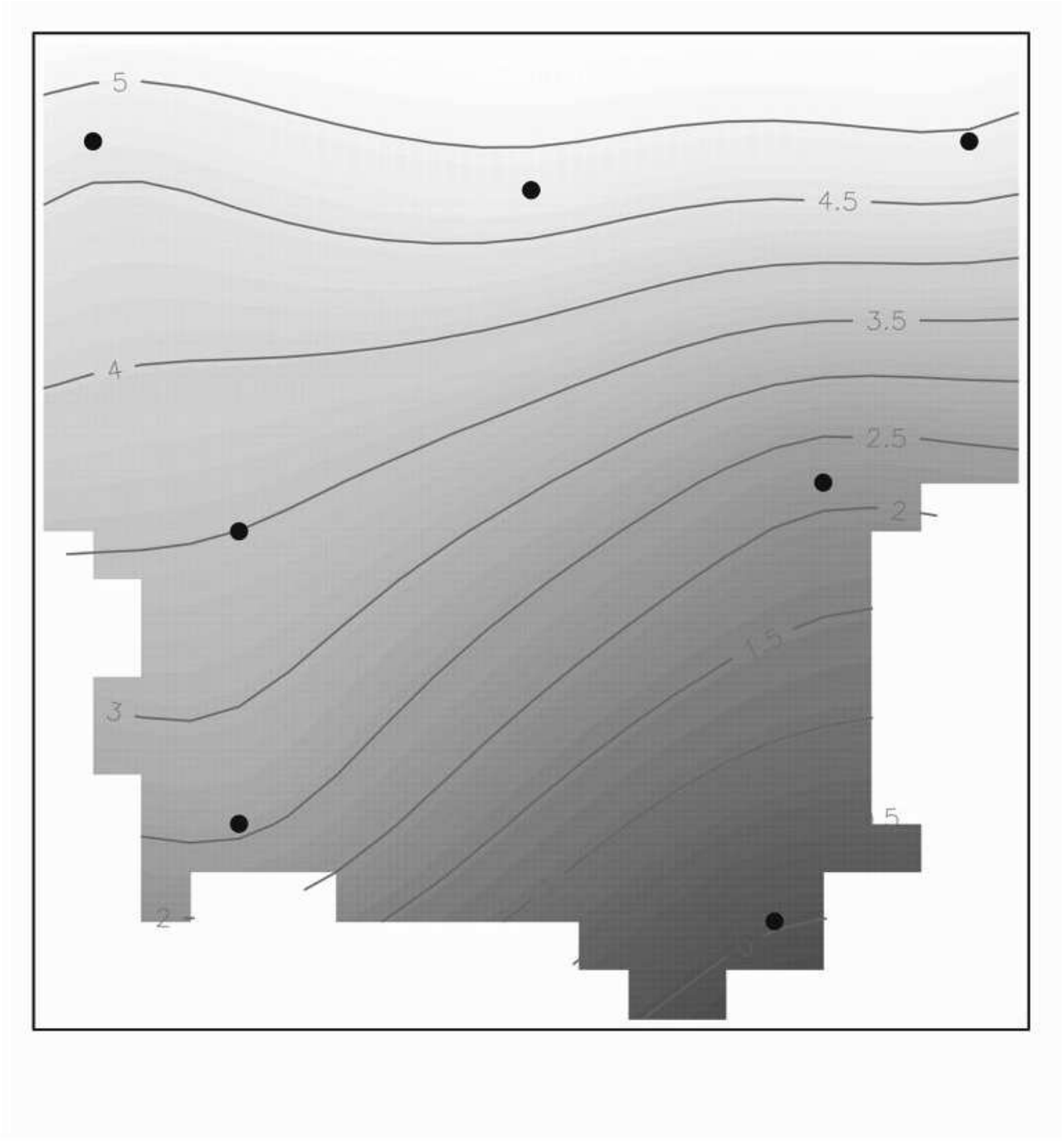}
 \end{center}
 \caption{\footnotesize Predicted fields for the best designs found. Left: Pareto-based algorithm; right: direct optimization of EK.}\label{fig:kpred_bestalpha}
\end{figure}

In Figure \ref{fig:BestDesignMUMM} we plot the corrected kriging variance for the designs obtained by the method on section \ref{S:SA-Pareto} (left) and by direct optimization of the Empirical Kriging criterion (right), overlaid with the corresponding optimal designs (indicated by the black dots).
We can see that while the Pareto-optimal design distributes the sampling points along the boundary of the region of analysis, the EK-optimal design contains several points in the interior of the design space, one at a considerable distance of the region boundary, and is able to keep the corrected kriging variance at lower levels $EK(\xi^\star) = 0.614$ versus $EK(\xi_{\SP}^\star) = 0.791$ (a space-filling design only gives $0.926$). Again, our Pareto-optimal yields a prediction error that was found to be better than $99\%$ of 10000 randomly generated sets of 7-point designs. Our sequential algorithm yielded another improvement to an EK-value of 0.761, albeit requiring 4 iterations with a total of 28 evaluations of the EK-criterion.

\begin{figure}
\begin{center}
 \includegraphics[bb= 0 60 369 390, width=.49\linewidth]{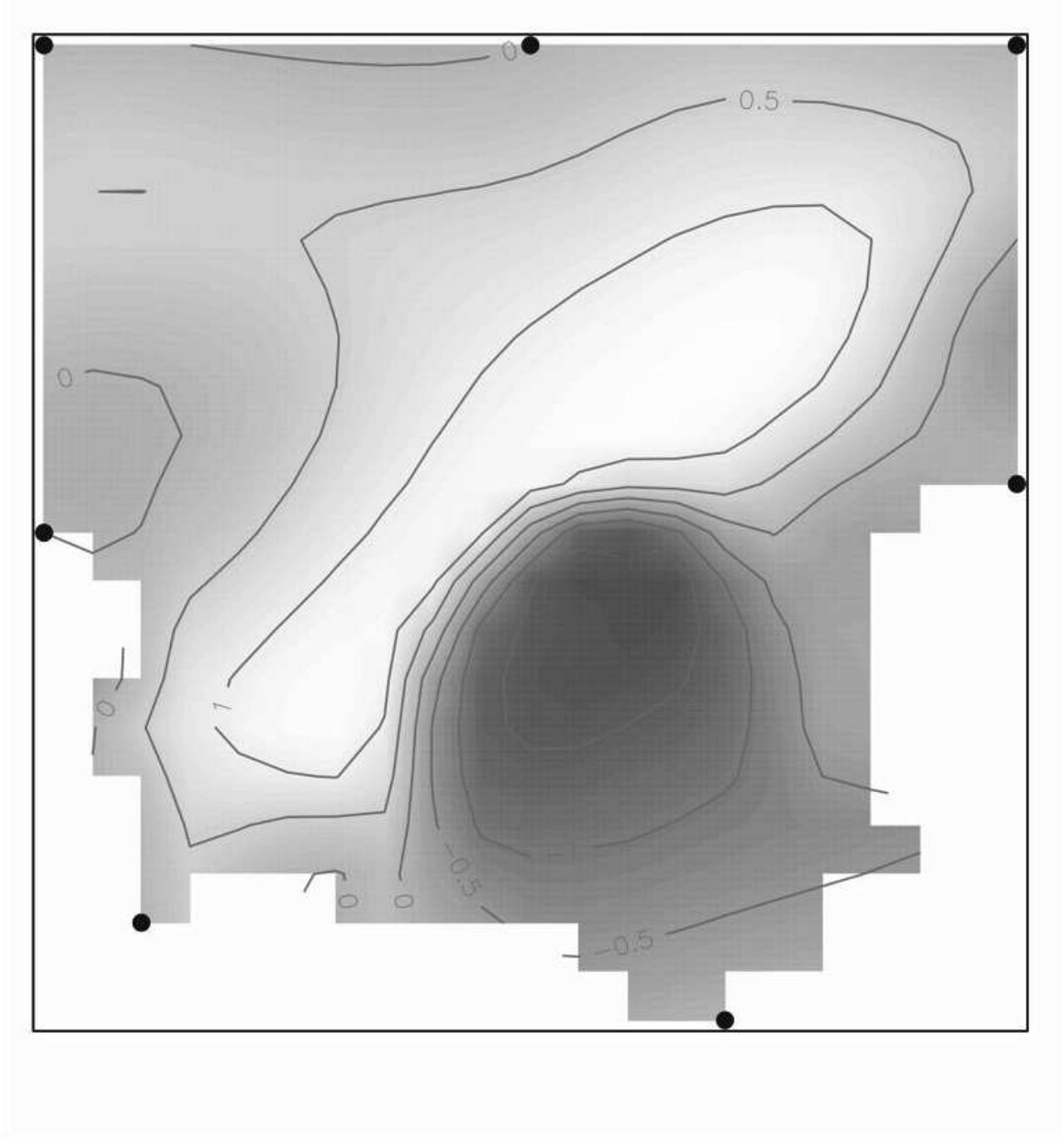}\includegraphics[bb= 0 60 369 390, width=.49\linewidth]{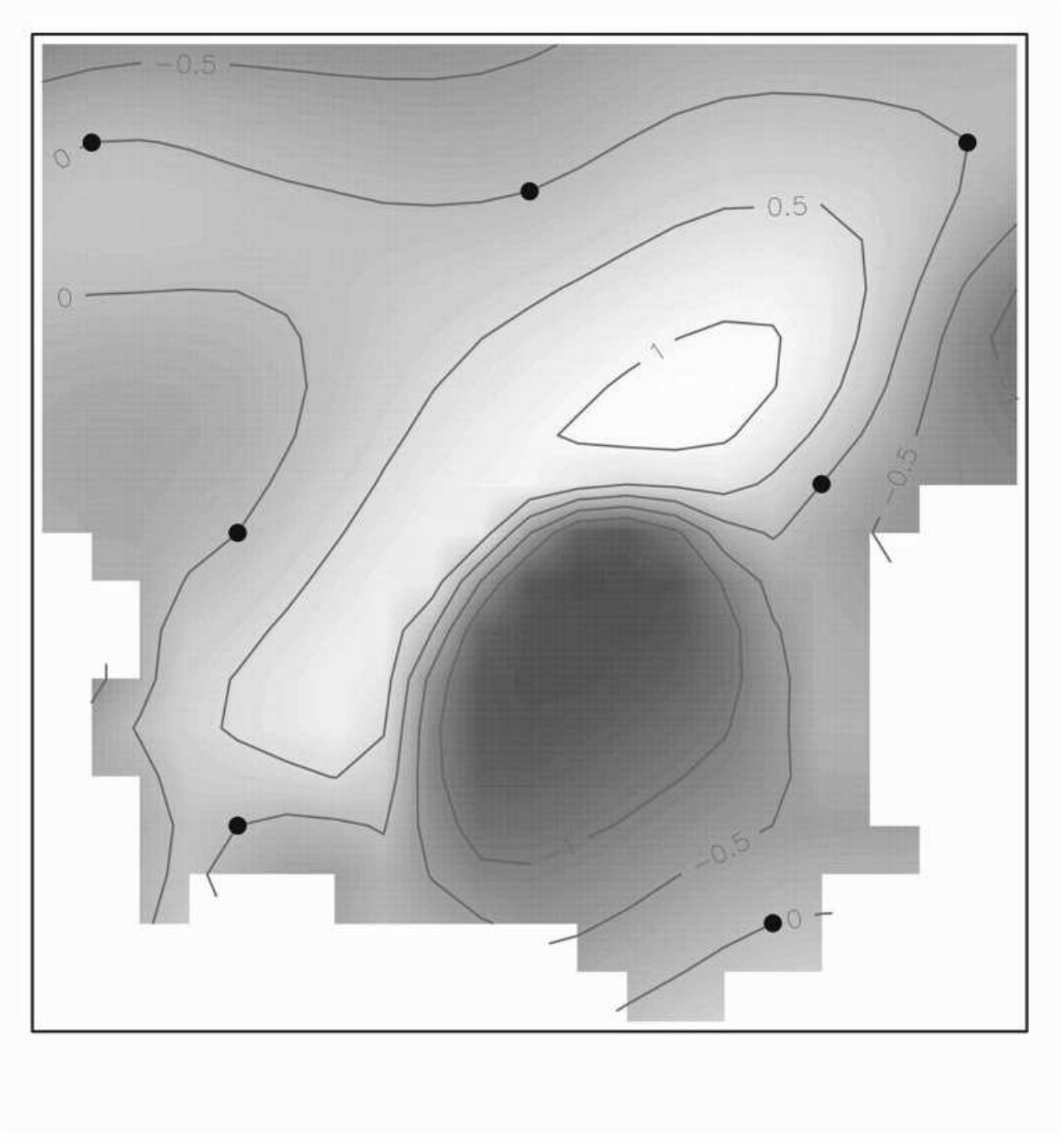}
 \end{center}
 \caption{\footnotesize prediction residuals fields for the best designs found. Left: Pareto-based algorithm; right: direct optimization of EK.}\label{fig:resids_bestalpha}
\end{figure}

Figures \ref{fig:kpred_bestalpha} and \ref{fig:resids_bestalpha} compare the fields predicted using these two designs, and the prediction residuals, respectively. Although the overall shape of the field is well reconstructed, being similar for both designs, inspection of the residual fields shows that the Pareto design leads to stronger deviations. In particular, both models fail to predict the high values of the field in the South-East small region, whose correlation structure strongly departs from the smoother variation in the open sea region, invalidating the predictions of the kriging variance. Note that these errors  are strong even for the EK optimal design, where a design point is located near that region.

Another  factor that may be affecting performance of the predictors in this region is related to the fact that the region of analysis is not convex, and thus the use of a covariance model based on simple Euclidean distance, like the Mat\'ern model, cannot capture the internal structure of the water mass, which is confined by the region bathymetry.

\section{MUMM example with $\rho=\frac{5}{2}$}\label{S:MUMMExample52}

\begin{figure}
\begin{center}
 \includegraphics[bb= 0 45 482 485, width=.49\linewidth]{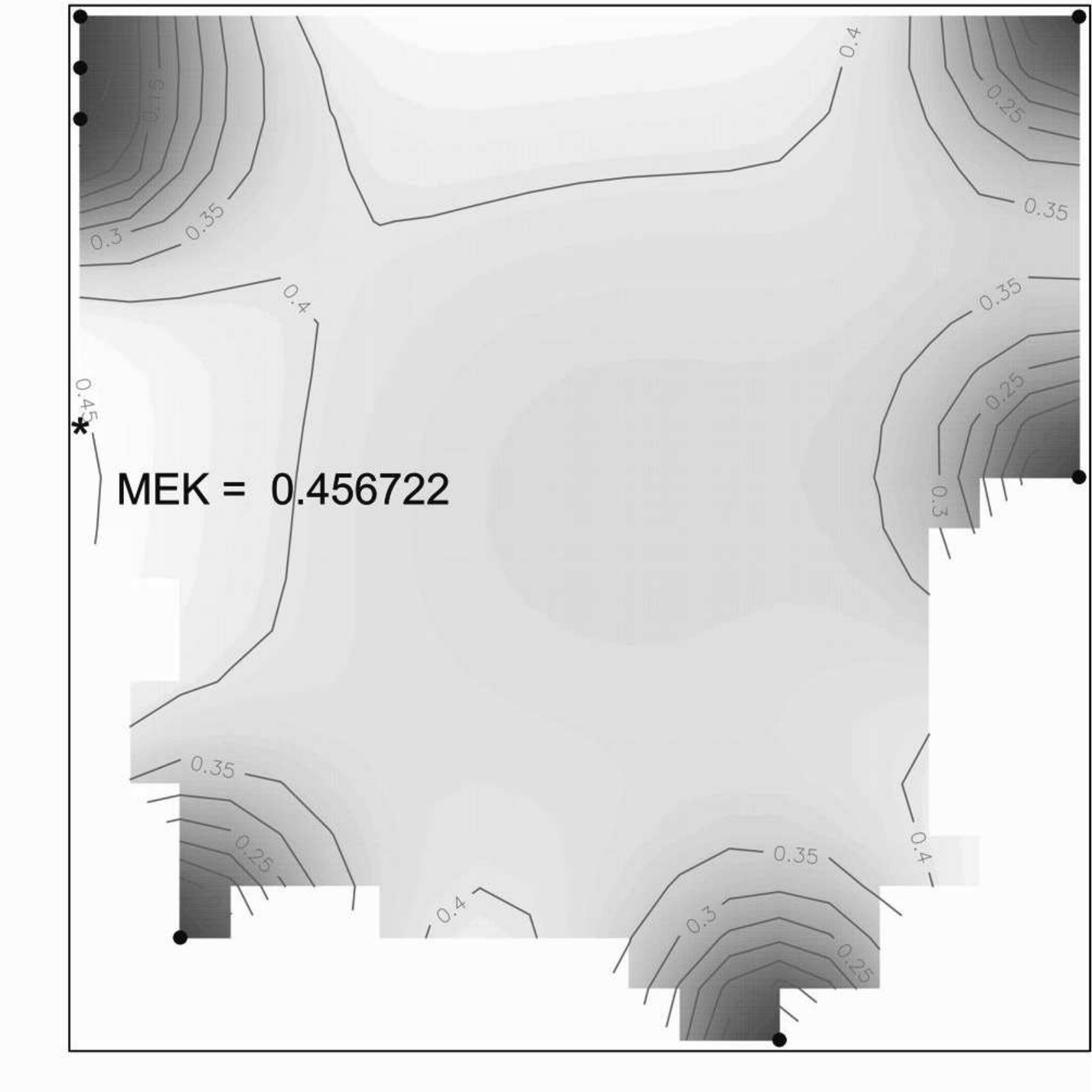}\includegraphics[bb= 0 45 482 485, width=.49\linewidth]{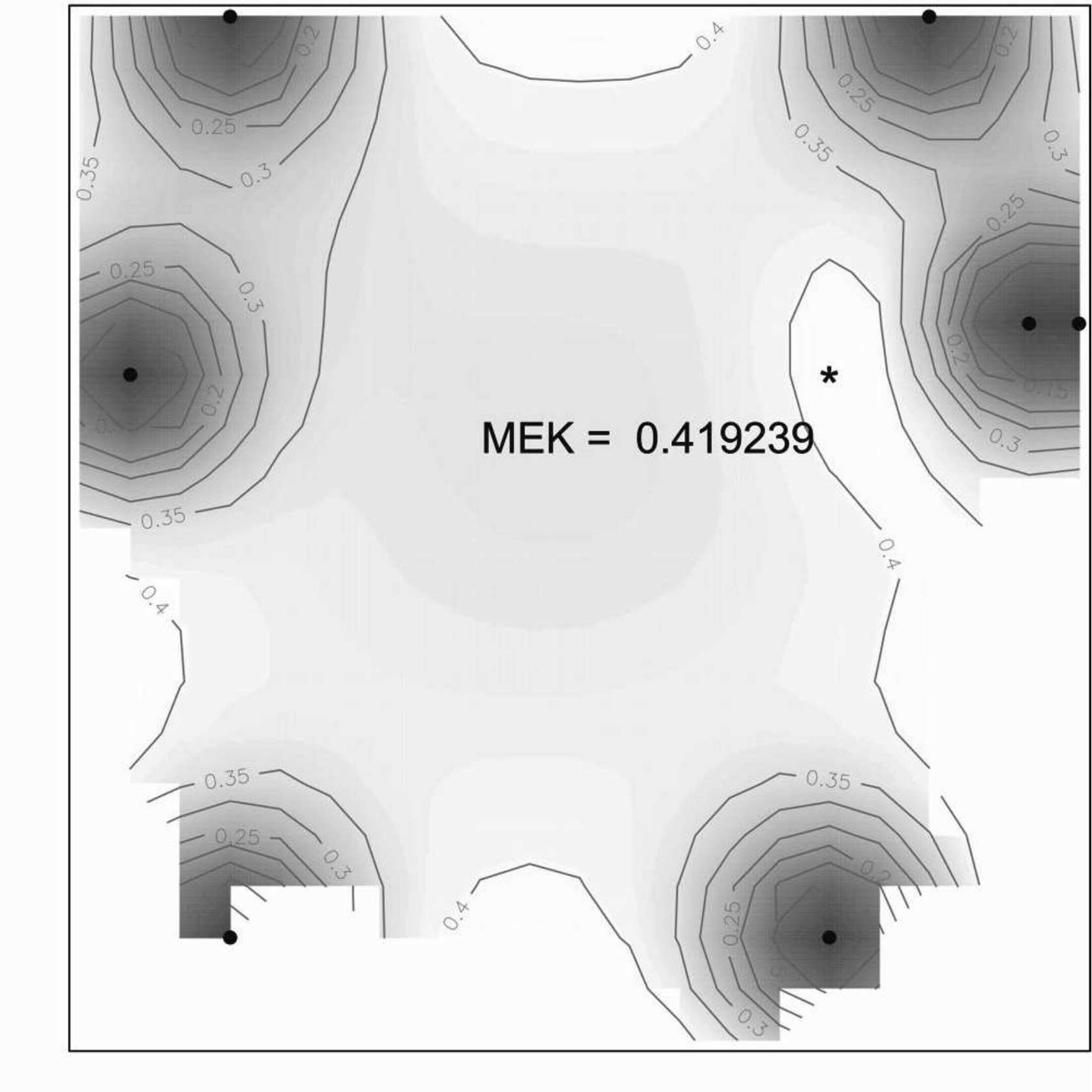}
\end{center}
\caption{\footnotesize Corrected Kriging variance over field of analysis for the best designs found. Left: Pareto-based algorithm; right: direct optimization of EK.}\label{fig:BestDesignMUMM52}
\end{figure}

A detailed analysis of the dataset used in \ref{S:MUMMExample} gave the following ML estimates for the covariance parameters:
\[\sigma^2 = 0.2927 \qquad \rho = 0.9826 \qquad \gamma = 2.664 \qquad \psi_{A}=1.104 \qquad \psi_{R}=1.328
\]
where $\psi_{A}$ and $\psi_{R}$ are anisotropy angle and ratio. OLS estimation of the covariance parameters gave
\[\sigma^2 = 0.6714 \qquad \rho = 1.1206 \qquad \gamma = 2.233
\]
On this evidence we fixed the smoothness parameter at $\gamma = 5/2$ moreover the model fit was improved with this parameter setting. The trend then was estimated as
\[
\eta(x_1,x_2, \beta) = -1.6242 -0.0639 x_1 +0.2338 x_2\enspace ,
\]
$\sigma^2 = 0.3098$, and range parameter $\rho = 1.0682$ and fixed $\gamma = 5/2$ gives

$c(x,x^\prime,\rho)=(1+\frac{\|x-x'\|}{\rho}+\frac{\|x-x'\|^{2}}{3\rho^{2}})\,\exp(-\|x-x'\|/\rho)$.

The  $7$ point Pareto-optimal design $\xi_{\SP}^\star$ for this model in the region of analysis has again been found by the method  presented in section \ref{S:SA-Pareto}, where again 6 distinct points were identified on the convex hull of the Pareto-surface. The  parameters of the SA algorithm were set as in Example 1, that was started from a random initialization.
The minimal Empirical Kriging variance was identified for $\alpha \in [0.6;0.8]$, indicating the importance of a good fit to the trend term in this case.

\begin{figure}
\begin{center}
 \includegraphics[bb= 0 45 482 485, width=.49\linewidth]{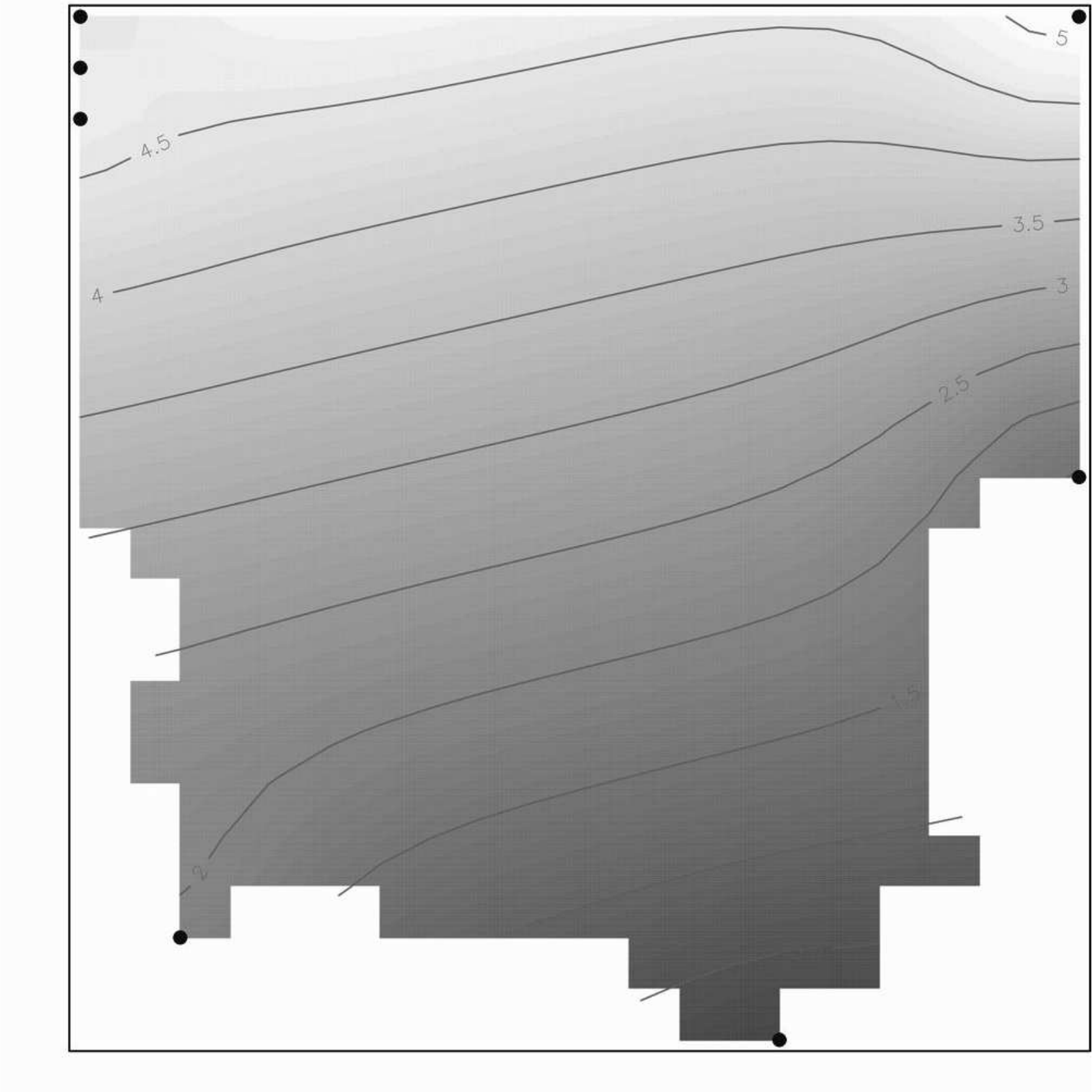}\includegraphics[bb= 0 45 482 485, width=.49\linewidth]{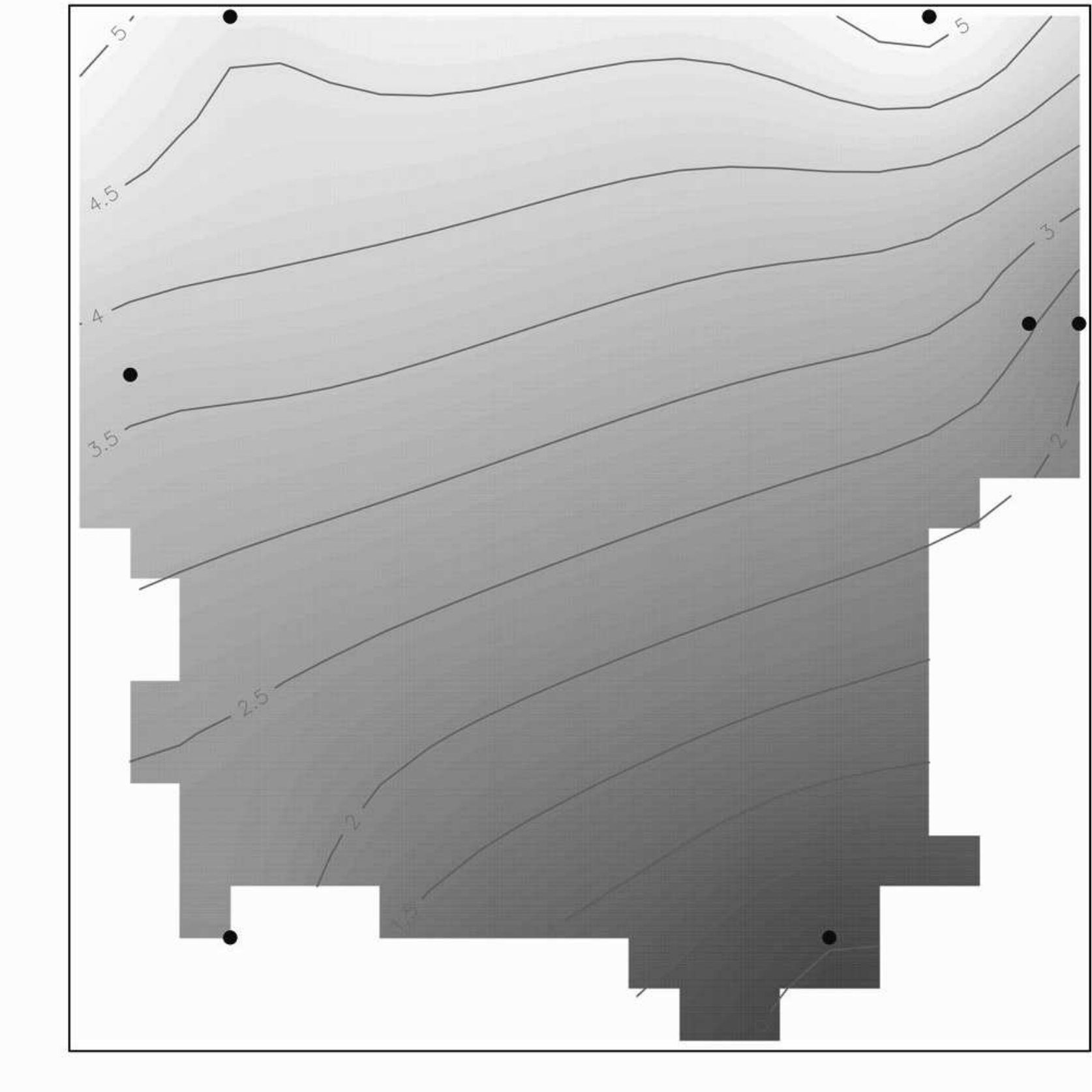}
 \end{center}
 \caption{\footnotesize Predicted fields for the best designs found. Left: Pareto-based algorithm; right: direct optimization of EK.}\label{fig:kpred_bestalpha52}
\end{figure}

In Figure \ref{fig:BestDesignMUMM52} we plot the corrected kriging variance for the designs obtained by the method on section \ref{S:SA-Pareto} (left) and by direct optimization of the Empirical Kriging criterion (right), overlaid with the corresponding optimal designs (indicated by the black dots).
We can see that while the Pareto-optimal design distributes the sampling points along the boundary of the region of analysis and tends to have one multiple sampling point, the EK-optimal design contains several points in the interior of the design space, one at a considerable distance of the region boundary, and is able to keep the corrected kriging variance at lower levels $EK(\xi^\star) = 0.4192$ versus $EK(\xi_{\SP}^\star) = 0.4567$ (a space-filling design (minimax) only gives $0.8508$ and a coffeehouse design $6.7462$). In this example the EK-efficiency of the Pareto-optimal design is 92\% whereas the EK efficiencies of minimax- and coffeehouse-design are only 49\% and 6\% which is rather poor.

\begin{figure}
\begin{center}
 \includegraphics[bb= 0 45 482 485, width=.49\linewidth]{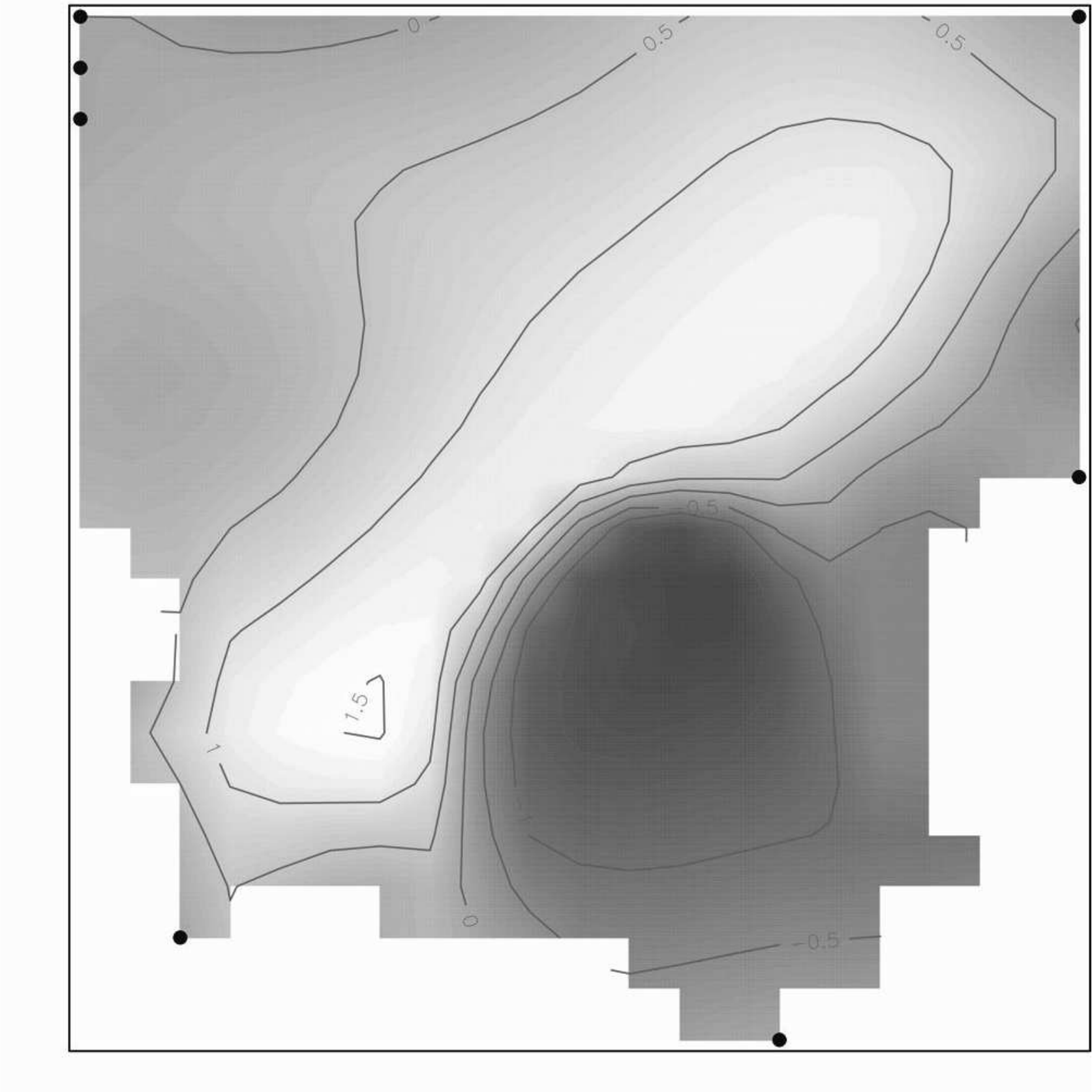}\includegraphics[bb= 0 45 482 485, width=.49\linewidth]{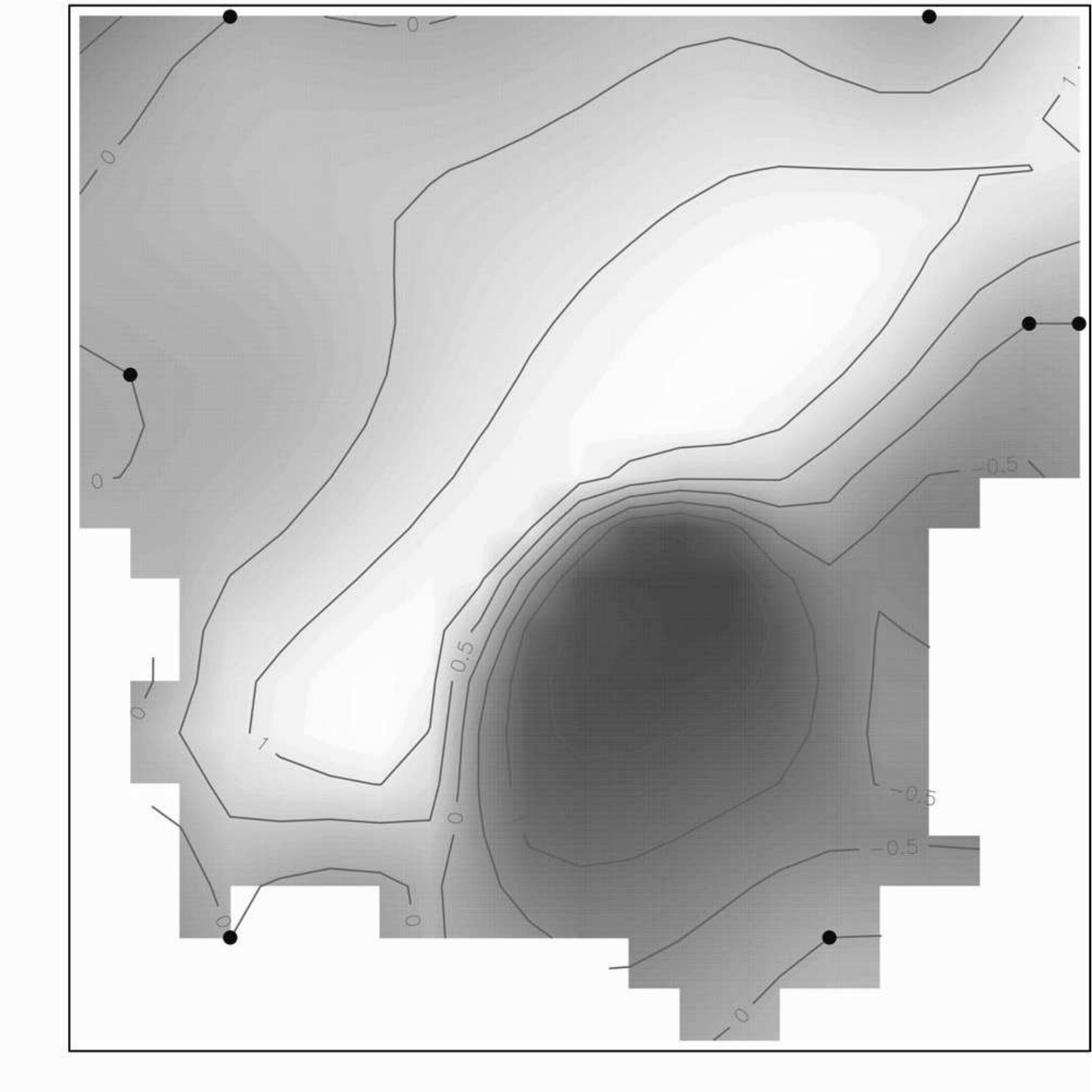}
 \end{center}
 \caption{\footnotesize prediction residuals fields for the best designs found. Left: Pareto-based algorithm; right: direct optimization of EK.}\label{fig:resids_bestalpha52}
\end{figure}

Figures \ref{fig:kpred_bestalpha52} and \ref{fig:resids_bestalpha52} compare the fields predicted using Pareto optimal and EK optimal designs, and the prediction residuals, respectively. Although the overall shape of the field is well reconstructed, being similar for both designs, inspection of the residual fields shows that the Pareto design leads to stronger deviations. In particular, both models fail to predict the high values of the field in the South-East small region, whose correlation structure strongly departs from the smoother variation in the open sea region, invalidating the predictions of the kriging variance. Note that these errors  are strong even for the EK optimal design, where a design point is located near that region.

\section{Conclusions}
\label{S:CONCL}
This paper proposes methods for identification of designs quasi-optimal for the corrected kriging variance in the context of prediction of spatial Gaussian fields. The criterion, also known as Empirical Kriging (EK) criterion, that takes into account the increased variance do to limited accuracy of the estimates of the covariance of the Gaussian process, is especially important when  this uncertainty is expected to make respective designs less space-filling.

Two methods are presented, both based on using the estimation criteria for the process parameters (related to the trend and to the covariance of the random term), that are to be simultaneously optimised, as surrogate criteria for the EK-minimisation, They offer significant increased efficiency compared to direct optimisation of the corrected kriging variance, by limiting  the evaluation of the numerically expensive EK-criterion to the Pareto-front of the two criteria. They differ significantly on how the Pareto-surface is determined. While one of the methods relies on the use of stochastic optimisation (SA) to sample the Pareto-front by optimising distinct convex combinations of the two criteria, the second is deterministic, and iteratively approaches this surface. They have characteristics that are dual in some sense: while in the first the number of sampled points of the Pareto surface is fixed by design (by the number of convex combinations that are optimised), in the second the number of evaluations of the EK criterion is not fixed in advance. The price payed for this controlled complexity is a potentially poorer sampling of the Pareto surface, leading eventually to a larger error of the chosen design.

The paper illustrates the two methods both in a simple simulated model, and also to a real oceanography data set. The results obtained show the validity of the approach underlying the two algorithms, that are able to identify designs that are close to optimal efficiency, and prediction variances that may be significantly lower than it would be possible using standard space filling designs. Of course, as we remark in the introductory sections of the paper, efforts to optimise the Empirical Kriging criterion should be limited to those situations where cost of observations is  large and the impact of the estimation of the covariance parameters cannot be neglected. In these cases, the methods proposed here offer a cost-effective alternative to the prohibitive direct optimisation of the relevant EK-criterion.

\section*{Acknowledgements}

The authors express their gratitude to Petra Vogl for computations and Jean-Marc F\'edou, Bertrand Gauthier, Gilles Menez, \'Eric Thierry and Milan Stehl\'{\i}k for discussions.

\section*{Appendix: Information matrix and empirical kriging variance for the Mat\'ern covariance function}

We analyze the model $\,Y\left( x\right) = f^T(x)\beta+\varepsilon(x)\,$ with Gaussian $\varepsilon \left( x\right)$ with zero mean and  Mat\'ern covariance (cf. eg. \cite{Stein99})
\[\Ex[\varepsilon \left( x_{i}\right)\varepsilon \left( x_{j}\right)]=\sigma^2 c(x_{i},x_{j},\nu)=\sigma^2 (C_{\nu})_{ij} = \sigma^2 \frac{\left(\frac{2d_{ij}\sqrt{\gamma}}{\rho}\right)^{\gamma}K_{\gamma}\left(\frac{2d_{ij}\sqrt{\gamma}}{\rho}\right)}{\Gamma(\gamma)2^{\gamma-1}}
\]
where $d_{ij}=|x_{i}-x_{j}|$, $\Gamma(\cdot)$ is the gamma function, $K_{\gamma}(\cdot)$ is the modified Bessel function of the second kind with order $\gamma$ and $\nu=(\rho,\gamma)^{T}$ are the non-negative covariance parameters.

\subsection*{A.1 Information matrix for the variance-covariance parameters}

Let $\theta=(\sigma^{2},\nu^{T})^{T}$ be the variance-covariance parameters of the  Mat\'ern covariance function. Then the information matrix for $\theta$ and a design $\xi=(x_{1},\ldots,x_{n})$ is given by (\ref{Mtheta}). That means, we have to compute the derivatives
\begin{equation*}
\frac{\partial (C_{\nu})_{ij}}{\partial \nu}=\left(\begin{array}{cc}
        \frac{\partial (C_{\nu})_{ij}}{\partial \rho} & \frac{\partial (C_{\nu})_{ij}}{\partial \gamma}
        \end{array}
        \right)^{T}
\tag{A1}
\label{der}
\end{equation*}

with $C_{\nu}$ simplified to $(C_{\nu})_{ij} = \frac{2}{\Gamma(\gamma)} \left(\frac{d_{ij}\sqrt{\gamma}}{\rho}\right)^{\gamma} K_{\gamma}\left(\frac{2d_{ij}\sqrt{\gamma}}{\rho}\right)$.
\bigskip

\underline{\textbf{Derivative with respect to }$\rho$\textbf{:}}
\smallskip

The computation of the derivative with respect to $\rho$ is straightforward. We just have to apply the product rule using $\frac{\partial K_{\gamma}(z)}{\partial z} = -\frac{1}{2}\left(K_{\gamma-1}(z)+K_{\gamma+1}(z) \right)$ (see \cite{Abr_72}) and then apply the following Bessel function identity (see \cite{Wei_12}).
\begin{equation*}
K_{\gamma-1}(z)=K_{\gamma+1}(z)-\frac{2\gamma}{z}K_{\gamma}(z)\enspace .
\tag{A2}
\label{ident}
\end{equation*}
\begin{eqnarray}\nonumber 
\frac{\partial (C_{\nu})_{ij}}{\partial \rho} & = & -\frac{2 \gamma  \left(\frac{d_{ij} \sqrt{\gamma }}{\rho }\right)^{\gamma }K_{\gamma}\left(\frac{2d_{ij}\sqrt{\gamma}}{\rho}\right)}{\rho  \Gamma (\gamma )} + \frac{2 \left(\frac{d_{ij} \sqrt{\gamma }}{\rho }\right)^{\gamma +1} \left(K_{\gamma -1}\left(\frac{2 d_{ij} \sqrt{\gamma }}{\rho }\right)+K_{\gamma +1}\left(\frac{2 d_{ij} \sqrt{\gamma }}{\rho }\right)\right)}{\rho  \Gamma (\gamma )}= \\
\nonumber & = & \frac{4}{\rho \Gamma (\gamma )} \left(\frac{d_{ij} \sqrt{\gamma }}{\rho }\right)^{\gamma +1} K_{\gamma -1}\left(\frac{2 d_{ij} \sqrt{\gamma }}{\rho }\right)
\end{eqnarray}
\medskip

\underline{\textbf{Derivative with respect to the order }$\gamma$\textbf{:}}
\smallskip

The computation of the derivative with respect to $\gamma$ is more complicated. First we have to apply the product rule using the polygamma function of order $0$, $\psi ^{(0)}(\gamma )=\frac{\partial\Gamma}{\partial\gamma}(\gamma) \frac{1}{\Gamma(\gamma)}$. Finally we again have to apply the identity (\ref{ident}).
\begin{eqnarray}\nonumber
\frac{\partial (C_{\nu})_{ij}}{\partial \gamma} & = & \frac{2\left(\frac{d_{ij}\sqrt{\gamma}}{\rho}\right)^{\gamma}} {\Gamma(\gamma)} \left(\ln\left(\frac{d_{ij}\sqrt{\gamma}}{\rho}\right)+\frac{1}{2}\right) K_{\gamma}\left(\frac{2d_{ij}\sqrt{\gamma}}{\rho}\right) - \frac{2\left(\frac{d_{ij}\sqrt{\gamma}}{\rho}\right)^{\gamma}} {\Gamma(\gamma)} \cdot\psi^{(0)}(\gamma ) \cdot K_{\gamma}\left(\frac{2d_{ij}\sqrt{\gamma}}{\rho}\right) + \\
\nonumber & & + \frac{2\left(\frac{d_{ij}\sqrt{\gamma}}{\rho}\right)^{\gamma}} {\Gamma(\gamma)} \left(\frac{\partial K_{\gamma}}{\partial\gamma}\left(\frac{2 d_{ij} \sqrt{\gamma }}{\rho }\right) - \frac{d_{ij}}{2\sqrt{\gamma}\rho} \left(K_{\gamma-1}\left(\frac{2d_{ij}\sqrt{\gamma}}{\rho}\right) +K_{\gamma+1}\left(\frac{2d_{ij}\sqrt{\gamma}}{\rho}\right) \right)\right) = \\
\nonumber & = & \frac{2\left(\frac{d_{ij}\sqrt{\gamma}}{\rho}\right)^{\gamma}} {\Gamma(\gamma)} \left(\frac{\partial K_{\gamma}}{\partial\gamma}\left(\frac{2 d_{ij} \sqrt{\gamma }}{\rho }\right) +K_{\gamma }\left(\frac{2 d_{ij} \sqrt{\gamma }}{\rho }\right) \left(\ln \left(\frac{d_{ij} \sqrt{\gamma }}{\rho }\right)-\psi ^{(0)}(\gamma )\right) - \right. \\
\nonumber & & \left. -\frac{d_{ij}}{\sqrt{\gamma}\rho} K_{\gamma -1}\left(\frac{2 d_{ij} \sqrt{\gamma }}{\rho }\right)\right)
\end{eqnarray}
For the derivative of the modified Bessel function of the second kind we have to compute
\[
\frac{\partial K_{\gamma}}{\partial\gamma}(z) = \begin{cases} \frac{\pi\csc(\gamma\pi)}{2}\left(-2\cos(\gamma\pi)K_{\gamma}(z)-\ln\left(\frac{z}{2}\right)(I_{-\gamma}(z)+I_{\gamma}(z)) + \right. & \\
\left.+\sum_{k=0}^{\infty}\frac{1}{k!} \left(\frac{\psi^{(0)}(k-\gamma+1)}{\Gamma(k-\gamma+1)} \left(\frac{z}{2}\right)^{2k-\gamma}+\frac{\psi^{(0)}(k+\gamma+1)}{\Gamma(k+\gamma+1)} \left(\frac{z}{2}\right)^{2k+\gamma}\right)\right) & \textrm{for }\gamma\in\hspace{-0.3cm}/\;{\rm I}\!{\rm N} \vspace{0,5cm}\\
\frac{\gamma!}{2}\left(\frac{z}{2}\right)^{-\gamma} \sum_{k=0}^{\gamma-1}\frac{K_{k}(z)}{(\gamma-k)k!}\left(\frac{z}{2}\right)^{k} & \textrm{for }\gamma\in{\rm I}\!{\rm N}
\end{cases}
\]
where $I_{\gamma}(\cdot)$ is the modified Bessel function of the first kind with order $\gamma$ (see \cite{Abr_72}, \cite{Wei_12}).

\subsection*{A.2 Empirical kriging variance for the Matern covariance function}

In order to find EK-optimal designs we have to minimize the design space maximum of the corrected kriging variance (see equation (\ref{EK2}))
\begin{equation}
\textrm{Var}\left[\hat{Y}(x)\right]+\textrm{tr}\left(V_{\nu}\textrm{Var}\left[\frac{\partial\hat{Y}(x)}{\partial\nu}\right]\right)\qquad x\in\SX\enspace .
\tag{A3}
\label{ekv}
\end{equation}
Here $\hat{Y}(x)=v^{T}Y(x)$ is the kriging prediction for design $\xi$ at point $x\in\SX$. Let $\sigma^{2}c_{n}$ be the vector of covariances between $x$ and design points $\xi$, then we have $v^{T}=c_{n}^{T}C_{\nu}^{-1}+(x-c_{n}^{T}C_{\nu}^{-1}X)(X^{T}C_{\nu}^{-1}X)^{-1}X^{T}C_{\nu}^{-1}$ where $X=(f(x_1),\dots, f(x_n))^T$ is the design matrix for the given model.

Then the classic kriging variance is $\textrm{Var}[\hat{Y}(x)]=\sigma^{2}(1+v^{T}C_{\nu}v-2v^{T}c_{n})$ and the correction term in (\ref{ekv}) equals to $\sigma^{2}\cdot\textrm{tr}\left(V_{\nu}\cdot\frac{\partial v^{T}}{\partial\nu}C_{\nu}\frac{\partial v}{\partial\nu^{T}}\right)$ where $V_{\nu}$ and
\[
\frac{\partial v^{T}}{\partial\nu}=\left(\frac{\partial c_{n}^{T}}{\partial\nu}-v^{T}\frac{\partial C_{\nu}}{\partial\nu}\right)C_{\nu}^{-1}\left(I-X(X^{T}C_{\nu}^{-1}X)^{-1}X^{T}C_{\nu}^{-1}\right)\enspace
\]
again depends on the derivatives (\ref{der}) of the Matern covariance function.

{}

\end{document}